\documentclass[useAMS,usenatbib,onecolumn,usegraphicx]{mn2e}
\usepackage{color}
\usepackage{amssymb}
\usepackage{natbib}

\title[
Instability of the nonlinearly-saturated MR mode
]
{
Resonant instability of the nonlinearly-saturated magnetorotational mode in thin Keplerian discs
}
\author[Yuri M. Shtemler, Michael Mond, and Edward Liverts]
{Yuri M. Shtemler$^{1}$\thanks{E-mail:
shtemler@bgu.ac.il; mond@bgu.ac.il; eliverts@bgu.ac.il},  Michael Mond$^{1}$, and  Edward  Liverts$^{1}$\\
$^{1}$Department of Mechanical Engineering,  Ben-Gurion
University of the Negev,  P.O. Box 653, Beer-Sheva 84105,
Israel
}
\begin{document}

\date{Accepted ---. Received ----; in original form ----}

\pagerange{\pageref{firstpage}--\pageref{lastpage}} \pubyear{}

\maketitle

\label{firstpage}

\begin{abstract}
The magneto-rotational decay  instability (MRDI)  of  thin Keplerian discs threaded by poloidal magnetic fields is introduced and studied. The linear magnetohydrodynamic problem decouples into eigenvalue problems for in-plane slow- and fast- Alfv\'en-Coriolis (AC), and vertical magnetosonic (MS) eigenmodes. The   magnetorotational instability (MRI) is composed of a discrete number of unstable slow AC eigenmodes that is determined for each radius by the local beta.  In the vicinity of the first beta threshold a parent  MRI eigenmode together with a stable AC eigenmode (either slow or fast) and a stable MS eigenmode form a resonant triad. The  three-wave MRDI  relies on the  nonlinear saturation of the parent MRI mode and
 the exponential growth of two daughter linearly stable waves, slow-AC and MS modes  with an effective growth rate that is comparable to that  of the parent MRI.
If, however, the role of the
AC daughter wave is played by a stable fast mode,  all three modes remain bounded.
\end{abstract}

\section{Introduction}
The linear magnetohydrodynamical (MHD) stability of thin Keplerian discs has been a focus of intensive investigation over the last two decades pertaining to the basic astrophysical problem of the origin of turbulence and enhanced angular momentum transfer in such discs. The original linear theory of the magnetorotational instability (MRI) was developed for infinite axially-uniform cylinders by [\cite{Velikhov 1959};  \cite{Chandrasekhar 1960}] who discovered the destabilizing effect of an axial magnetic field on Couette flow. That phenomenon has been conjectured to play the main role in the transition to turbulence in astrophysical discs as well as in the subsequent enhanced angular momentum transfer by [\cite{Balbus and Hawley 1991}], who considered a "cylindrical" model of Keplerian discs. That ground breaking work has been followed by numerous analytical as well as numerical investigations under a wide range of conditions and applications. Employing asymptotic methods that are relevant to thin disc geometry,  linear stability studies demonstrated the stabilizing effects of the finite disc's thickness
as well as the discrete nature of the MRI spectrum
[
\cite{Balbus and Hawley 1991}, \cite{Gammie and Balbus 1994},
\cite{Coppi and Keyes 2003}, \cite{Liverts and Mond 2009}, \cite{Shtemler et al. 2011}]. In addition, a weakly nonlinear analysis was employed in order to demonstrate the dissipative saturation of the MRI
[\cite{Knobloch and Julien 2005},
 \cite{Umurhan et al. 2007}]. Thus, \cite{Knobloch and Julien 2005}
 have described analytically the nonlinear saturation of the MRI in a straight infinite vertically uniform channel with solid boundaries, a configuration typical for laboratory experiments rather than astrophysical conditions. \cite{Knobloch and Julien 2005}  considered a developed stage of the MRI, far from its threshold and showed that as the presence of the solid boundaries supports radial pressure gradients, the latter act together with the viscous as well as Ohmic dissipation in order to modify the rotation shear that feeds the instability and thus saturating it. In a complementary work, \cite{Umurhan et al. 2007} performed a weakly nonlinear analysis of the MRI close to marginality in configurations similar to \cite{Knobloch and Julien 2005}  and showed that the MRI saturates due to dissipative effects to levels that scale with the square root of the magnetic Prandtl number. As distinct from the last two cited works, \cite{Liverts et al. 2012a} and \cite{Liverts et al. 2012b} proposed the novel dissipationless mechanism for the saturation of the MRI that is realized through the nonlinear interaction of a MRI mode with an MRI-driven magnetosonic (MS) wave. The weakly-nonlinear model by \cite{Liverts et al. 2012a} and \cite{Liverts et al. 2012b} is based on a true thin disc geometry that is characterized by vertically localized number density. Such a mechanism leads to the saturation of the MRI [\cite{Liverts et al. 2012a} and \cite{Liverts et al. 2012b}] in the form of constant-amplitude bursty  nonlinear oscillation. Therefore, the saturation mechanism raised questions as to the efficiency of the MRI to directly generate significant levels of turbulence in thin accretion discs. The current work suggests that the MRI may still provide a viable mechanism for generating MHD turbulence in thin astrophysical discs, however it may do so in a much more indirect, subtle, and intricate way that has been thought so far.

True thin disc geometry is of principal significance for both linear and weakly-nonlinear stability analysis. Indeed, \cite{Balbus and Hawley 1991} solved the axially-uniform linear-stability problem and found the system to be unstable formally for all values of the plasma beta and for a finite domain of the axial wavenumber (including zero wavenumber). Recognizing that such consideration is rather applicable to infinite cylinders than to thin discs, a phenomenological bound for the critical axial wavelength was invoked by assuming that the latter shouldn't exceed the effective thickness of the disc. Using the dispersion relation for the cylindrical model a threshold value for the beta needed for instability is thus obtained. In contrast to the continuous nature of the MRI spectrum that is obtained from the cylindrical disc's model [\cite{Balbus and Hawley 1991}], the non-local thin-disc approximations in
\cite{Gammie and Balbus 1994},
 \cite{Liverts and Mond 2009}, \cite{Latter et al. 2010} and \cite{Shtemler et al. 2011}  results in a discrete spectrum. This is a direct result of the
  zero
   boundary conditions for the perturbed magnetic fields
   far away from the disc mid-plain.
    Notwithstanding the different nature of the two spectra, they are in fair correspondence for large values of the beta parameter for which the discrete spectrum becomes dense and tends to the continuous one in the limit of infinite beta [\cite{Shtemler et al. 2011}]. It should be stressed however, that the threshold beta is rigorously found without any ad hoc hypothesis within the non-local analysis.

The current study presents a generalization of the weakly nonlinear analysis of the dissipationless saturation of the MRI in thin discs so that resonant interactions between the various modes in the systems are also  taken into account [\cite{Shtemler et al. 2013}]. The weakly nonlinear analysis is characterized by two  dimensionless parameters: the plasma beta, $\beta$ , and disc's aspect ratio, $\epsilon$  (or, alternatively, the Mach number, $M \sim 1/\epsilon$).
  For any fixed beta the asymptotic thin-disc approximation allows to
  simplify the equilibrium  problem
\cite{Ogilvie 1997},
as well as
  the linear and   nonlinear problems of the disc stability  [\cite{Regev 1983},
   \cite{Ogilvie 1998},
   \cite{Umurhan et al. 2006}, \cite{Shtemler et al. 2007}, (2009), (2011)].
   A corresponding asymptotic procedure in   small  $\epsilon$  using additionally  a simplifying assumption  that  radial wavelengths that characterize the perturbations are  much larger the disc thickness results in eliminating the radial derivatives of the perturbed variables.
   This
    leads to a reduced system of equations
    that parametrically depends on the radial coordinate  (other versions of thin-disc approximation allow notice to eliminate the  perturbations to have a radial wavelength comparable to the disc thickness [\cite{Ogilvie 1998}]).
     Thus, assuming axially isothermal thin Keplerian discs, the linear MHD problem decouples into two in-plane Alfv\'en-Coriolis (AC) modes and stable vertical MS mode. One of the AC modes may become unstable (this is the familiar MRI) while the other one is stable for all values of the relevant parameters. The linear  problem for the MRI mode was solved by \cite{Liverts and Mond 2009} by employing the Wentzel-Kramers-Brillouin (WKB) approximation. Formally restricted to large beta values,  the WKB solution can be shown to be effective also for $\beta \sim 1$. Alternatively, in thin disc approximation a full analytical solution of the linear MRI problem is found for arbitrary beta values by approximating the axially-isothermal Gaussian profile of the background density by a hyperbolic function [see
     \cite{Gammie and Balbus 1994}, see also  \cite{Latter et al. 2010} and \cite{Shtemler et al. 2011},
     as well as Section 3.1 in the present study]. In any case, axial stratification of the steady-state density (be it Gaussian or hyperbolic) is a crucial feature for facilitating the dissipationless mechanism that is proposed in the current work. This may be easily inferred from the analysis of the AC modes which propagate along the magnetic field in rotating isothermal axially-uniform infinite plasmas [ \cite{Landau and Lifshitz 1984}]. Such an analysis formally reduces the nonlinear model to the linear theory of the MRI due to the resulting constancy of the total perturbed pressure. Consequently, models that rely on axially uniform infinite discs, though successful in describing the linear stage of the MRI in the large beta limit, lack the ability to capture weakly nonlinear effects and in particular the mechanism that gives rise to non-dissipative saturation. In some sense the analysis presented in [\cite{Landau and Lifshitz 1984}] is a predecessor of the channel modes notion [\cite{Goodman    and Xu 1994}], which for incompressible and axially uniform discs extends the validity of the linear MRI solutions into the nonlinear regime.

As was established in \cite{Liverts et al. 2012a}, (2012b), the MRI can be nonlinearly saturated by periodically transferring its energy to MS modes. The saturation occurs due to the vertical stratification of a disc and the physically plausible
 zero
  boundary conditions. Indeed, as demonstrated in \cite{Liverts et al. 2012a} and \cite{Liverts et al. 2012b}, the amplitude of the MRI mode is described by Duffing's equation: $d^2 A/dt^2=E_0 A+E_2 A^3+\dots$ (with $E_0 =\gamma^2$, $\gamma$ is the growth rate of the linear MRI mode), whose solution  exhibits cyclic saturation in time for $E_0>0$, $E_2<0$ [\cite{Guckenheimer and Holmes 1983}].  In that connection note the study by \cite{Arter 2009}  which used Duffing's equation in order to describe sawtooth oscillations in Tokamaks, note also that  qualitatively similar sawtooth oscillations in magnetic field has been obtained in  a spherically symmetric mean-field dynamo model for a proper normalized dynamo number \cite{Stefani  and Gerbeth 2005}. The specific form of the undamped Duffing equation follows from the double-zero degeneracy of the linear MRI problem at the threshold plasma beta. The derived Duffing equation differs from the classical Landau amplitude equation, $dA/dt=E_0 A+E_2|A|^2 A+\dots$ (with $E_0 =\gamma$), not only with respect to their forms, but also due to the origin of the nonlinear terms. While the cubic term in the classical Landau equation stems from self-distortion effects of the background mode, the cubic term in the Duffing equation is only formally reminiscent of the term arising due to self-excitation one. In fact it reflects the interaction of the background MRI mode with the MRI-driven MS one.

The solution of Duffing equation does indeed reflect the dissipationless saturation of the MRI.
Thus, for small amplitudes, the MRI behaves according to the linearized equation, however as the nonlinear term kicks in
the rate change of the amplitude increases as it grows,
 thus giving it a bursty appearance. Formally, the solution of the Duffing equation provides an accurate description of the evolution of the MRI near the stability threshold. However, direct numerical solutions of the full nonlinear set of reduced thin-disc MHD equations for initial conditions that are sufficiently far from the threshold suggest that non-dissipative oscillatory saturation of the instability occurs on realistic time scales (of the order of a few orbital times) also far away from the linear instability threshold [\cite{Liverts et al. 2012b}].
Adding enough dissipation removes the degeneracy of the linear MRI problem, which ultimately leads to the Landau-like amplitude equation [\cite{Umurhan et al. 2007}] instead of the Duffing equation [\cite{Liverts et al. 2012a} and \cite{Liverts et al. 2012b}]. However, under astrophysical conditions the non-dissipative saturation occurs at time scales that are much smaller than its dissipative counterpart, and the non-dissipative saturation is described by an intermediate asymptotics that precedes the start of dissipative processes.

In the present paper the weakly-nonlinear instability of Keplerian discs is investigated with respect to axisymmetric perturbations. A complete description of further development of the nonlinearly-saturated MRI is exceedingly difficult to achieve, yet a number of simple resonant interactions emerge as crucial building blocks for future studies. In this respect resonant triads are of particular importance because they are the lowest order nonlinear effects that provide the strongest possible interaction between the modes involved [\cite{Craik 1985}]. As will be demonstrated in the present paper resonant triads of stable MS and two AC modes (one of which is the first MRI mode) in thin discs in some principal respects significantly differs from the resonant triads of stable MS and both two AC modes in unbounded
stationary
 plasmas investigated in earlier study by [\cite{Galeev  and Oraevskii 1962-1963}, \cite{Sagdeev and Galeev 1969}].  The  magneto-rotational decay instability (MRDI)  introduced in the present study generalizes the physical mechanism of fundamental three-wave  instability in infinite, homogeneous and
stationary
 plasmas   to thin Keplerian discs. Such three-mode resonant interaction is particularly effective in thin discs due to their special geometry. Since the radial derivatives of the perturbed variables are neglected within
  the approximation adopted,
   the condition of resonant coupling of the radial wavenumbers is dropped from consideration [\cite{Shtemler et al. 2011}]. Consequently, the resonant conditions are reduced to a matching condition for the frequencies of the resonant participants. Those conditions are easily satisfied in turn due to the continuous nature of the spectrum of the MS modes
 proved for a hyperbolic profile of the background density
[\cite{Liverts et al. 2012a} and \cite{Liverts et al. 2012b}].
Extending thus the weakly nonlinear analysis to resonant triads entails a  surprising result. While the amplitude of the parent MRI saturates via periodical nonlinear oscillations just as in the non-resonant case, it is shown in the current work that the two daughter modes  (i.e. the linearly stable AC  and  MS modes) may grow exponentially on time scales of the inverse growth rate of the parent MRI eigenmode.

The paper is organized as follows. The physical model for thin Keplerina discs is presented in the next Section. Some results of linear stability analysis for thin discs are summarized in Section 3. Section 4 describes in details a non-resonant interaction between the MRI and forced-MS modes. The weakly nonlinear analysis of resonant triads is carried out in Section 5, where the resonant relations are derived and nonlinear coupling coefficients are calculated. Summary and discussion are given in Section 6.


\section{THE PHYSICAL MODEL FOR THIN KEPLERIAN DISCS.}

\subsection{Governing relations.}
Radially and axially stratified rotating thin plasma discs subject to poloidal magnetic fields are considered. The following characteristic values of the various physical variables   are used as scaling parameters:
\begin{equation}
V_*=C_{S*},\,\,\,
 H_*=V_*t_*,\,\,\,
\Phi_*=\frac{G\, M_c}{H_*}\equiv \frac{1}{\epsilon}M^2_* V^2_{*},\,\,\, m_*=m_i,\,\,\, n_*=n_i,\,\,\,
P_*=m_*n_*V_{*}^2,\,\,\,
j_*=\frac{c}{4\pi}\frac{B_*}{H_*},\,\,\,
E_*=\frac{V_*B_*}{c}.
\label{1}
\end{equation}
Here $V_*$, $C_{S*}=\sqrt{T_*/m_*}$, $t_*$,  $V_{K*}=r_*\Omega_*\equiv\sqrt{G\, M_c/r_*}$,   $H_*$,  $\Phi_*$,
   $m_*$,  $n_*$, $P_*$, $j_*$  and $E_*$
  are the characteristic dimensional  values: the fluid velocity, the sound velocity, time, the Keplerian velocity of the fluid, the semi-thickness of the disc,  the gravitational potential,  the ion mass and number density, the pressure, the electric current density and electric field;  $G$  is the gravitational constant; $M_c$ is the total mass of the central object; $c$ is the light speed; $T_*=T(r_*)$ and $B_*=B_z(r_*)$ are the  temperature and the  axial magnetic field at  the characteristic radius $r_*$   that belongs to the Keplerian portion of the disc.
  A hydrodynamic model of dissipationless quasi-neutral plasmas is characterized by three dimensional velocities, namely, the Keplerian rotation,   $V_{K*}$, the sound velocity,  $C_{S*}$, and the Alfv\'en velocity,  $U_{A*}=B_*/\sqrt{4\pi m_* n_*}$, which in turn produce two dimensionless parameters: the beta parameter, $\beta_*$ and the  Mach number $M_*$  (that equals to inverse disc aspect ratio [\cite{Frank et al. 2002}]):
 \begin{equation}
\beta_*=4\pi\frac{P_*}{B_{*}^2}\equiv \frac{C_{S*}^2}{U_{A*}^2},\,\,\, M_*=\frac{V_{K*}} {C_{S*}}\equiv \epsilon^{-1}
.\,\,\,
\label{2}
\end{equation}
%
%
%

  The resulting dimensionless dynamical equations are:
\begin{equation}
\frac{ D \bf {V} }{Dt} = -\frac{\nabla P}{n}+\frac{1}{\beta_* } \frac{\bf{j}
\times {\bf{B}}}{n}- \epsilon^{-1} M^2_*\nabla \Phi,\,\,\,\,\Phi(r,z)=-\frac{1}{(r^2+z^2)^{1/2}},
\label{3}
\end{equation}
\begin{equation}
\frac{\partial n} {\partial t}
               + \nabla \cdot(n {\bf{V}} ) =0,\\
\label{4}
\end{equation}
\begin{equation}
\frac{\partial {\bf{B} }} {\partial t}+
 \nabla \times
 {\bf{E}}=0,\,\,\, {\bf {E} }=  -  {\bf{V}} \times {\bf{B}},
\label{5}
\end{equation}
\begin{equation}
 {\bf {j} }= \nabla   \times {\bf{B}},\,\,\,\,\,\,\,\,\,\nabla \cdot {\bf{B}}=0,
\label{6}
\end{equation}
\begin{equation}
P =nT.
\label{7}
\end{equation}
Here $P$, $n$ and $T$ are the total plasma pressure, number density  and temperature; $\bf{V}$  is the plasma velocity. The disc is assumed to be vertically isothermal which implies that $\nabla P=\bar{C}^2_S\nabla n$, where the dimensionless sound speed is given by $\bar{C}^2_S =\partial P/\partial n\equiv \bar{T}(r)$,  $\bar{T}$  is the steady-state background temperature.
%
%
  Standard cylindrical coordinates
$\{r,\theta,z\}$   are adopted throughout the paper;
$t$ is time; $D/Dt=\partial/\partial
t+(\bf{V}\cdot\nabla)$  is the material derivative; $\Phi(r,z)$  is the
 gravitational potential due to the central body; $\bf{B}$, $\bf{j}$ and $\bf{E}$ are the magnetic field,    current density and
  electric field, respectively.
  Note that a preferred direction is tacitly defined here, namely, the positive direction of the z axis is chosen according to positive Keplerian rotation.

  Completing the formulation of the physical model, the
   following
     conditions are imposed at infinity:
\begin{equation}
B_r=B_\theta=0,\,\,\,n=0\,\,\,\mbox{for} \,\,\, z=\pm\infty.
\label{8}
\end{equation}
Note also that the mass flux as well as the axial momentum flux are zero at infinity.

 The smallness of   $\epsilon$   makes it useful to define the slow radial coordinate, and as the radial coordinate is a mere parameter to leading order in   $\epsilon$, it is convenient to replace the  independent variables by the following
 quantities:
\begin{equation}
\xi=\epsilon r\sim \epsilon^0,\,\,\,
\tau= \bar{\Omega}(\epsilon r)t\sim \epsilon^0,\,\,\,
\eta=\frac{z}{\bar{H}(\epsilon r)}\sim\epsilon^0.
\label{9}
\end{equation}

\subsection{The basic steady-state equilibrium Keplerian configuration.}
Employing the definitions above, the  equilibrium Keplerian configuration is described as follows to leading order in $\epsilon$:
$$
V_r=V_z= 0,\,\,\,V_\theta =\epsilon^{-1}\bar{V}(\xi)+O(\epsilon),\,\,\,
B_r=B_\theta= 0,\,\,\,B_z=\epsilon^0\bar{B}(\xi),\,\,\,
\,\,\,\,\,\,\,\,\,\,\,\,\,\,\,\,\,\,\,\,\,\,\,\,\,\,\,\,
\,\,\,\,\,\,\,\,\,\,\,\,\,\,\,\,\,\,\,\,\,\,\,\,\,\,\,\,
\,\,\,\,\,\,\,\,\,\,\,\,\,\,\,\,\,\,\,\,\,\,\,\,\,\,\,\,
\,\,\,\,\,\,\,\,\,\,\,\,\,\,\,\,\,\,\,\,\,\,\,\,\,\,\,\,
\,\,\,\,\,\,\,\,\,\,\,\,\,\,\,\,\,\,\,\,\,\,\,\,\,\,\,\,
\,\,\,\,\,\,\,\,\,\,\,\,\,\,\,\,\,\,\,\,\,\,\,\,\,\,\,\,
$$
\begin{equation}
\Phi(r,z)\equiv -\frac{\epsilon}{\sqrt{\xi^2+\epsilon^2z^2}}
=-\epsilon\bar{\Phi}(\xi)+\frac{1}{2}\epsilon^3\bar{C}_S^2(\xi)\bar{\psi}(\eta)
+O(\epsilon^5),\,\,\,\,
n(r,z)=\epsilon^0\bar{N}(\xi)\bar{n}(\eta)+O(\epsilon^2),
\label{10}
\end{equation}
where
$$
\bar{V}(\xi)=\xi\bar{\Omega}(\xi)\equiv \frac{1}{\sqrt{\xi}},\,\,\,
\bar{\Phi}(\xi)=\frac{1}{\xi},\,\,\,
\bar{H}(\xi)=\frac{\bar{C}_S(\xi)}{\bar{\Omega}(\xi)},\,\,\,
\bar{\psi}(\eta)=\eta^2,\,\,\,
\bar{n}(\eta)=\exp{(-\eta^2/2)}.
\,\,\,\,\,\,\,\,\,\,\,\,\,\,\,\,\,\,\,\,\,\,\,\,
\,\,\,\,\,\,\,\,\,\,\,\,\,\,\,\,\,\,\,\,\,\,\,\,\,\,\,\,
\,\,\,\,\,\,\,\,\,\,\,\,\,\,\,\,\,\,\,\,\,\,\,\,\,\,\,\,
\,\,\,\,\,\,\,\,\,\,\,\,\,\,\,\,\,\,\,\,\,\,\,\,\,\,\,\,\,\,\,\,\,\,\,\,\,\,\,\,\,
\,\,\,\,\,\,\,\,\,\,\,\,\,\,\,\,\,\,\,\,\,\,\,\,\,\,\,\,\,\,\,\,\,\,\,\,\,\,\,\,\,\,\,
\,\,\,\,\,\,\,\,\,\,\,\,\,\,\,\,\,\,\,\,\,\,\,\,\,\,\,\,
\,\,\,\,\,\,\,\,\,\,\,\,\,\,\,\,\,\,\,\,\,\,\,\,\,\,\,\,
$$
Note that Keplerian velocity,  $\bar{V}(\xi)$,  and Gaussian density,  $\bar{n}(\eta)$, are determined by the equilibrium conditions in the radial  and the axial directions, respectively, while  $\bar{C}_S(\xi)$,  $\bar{B}(\xi)$, $\bar{N}(\xi)$  are arbitrary functions, which form the local plasma beta
that parametrically depends on the radial variable, $\xi$:
\begin{equation}
\beta(\xi)=\beta_*\frac{\bar{N}(\xi)\bar{C}_S^2(\xi)}{\bar{B}^2(\xi)}.
\label{11}
\end{equation}

\subsection{Perturbed problem.}
Both steady-state equilibrium as well as the perturbed variables are scaled with well-defined powers of the small parameter  $\epsilon$:
\begin{equation}
F(r,z,t)=\epsilon^{\bar{S}}\bar{F}(\xi,\eta;\epsilon)+\epsilon^{S'} F'(\xi,\eta,\tau;\epsilon).
\label{12}
\end{equation}
Here  $F$ stands for any dependent variable, the bar and the prime denote equilibrium and perturbed variables, respectively, which are characterized by gauge functions $\epsilon^{\bar{S}}$  and  $\epsilon^{S'}$. The various values of $\bar{S}$  are determined above in Eqs. (10), while the values of $S'$ are assumed in a self-consistent way to be zero for all the dependent variables except for the axial component of the perturbed magnetic field for which  $S'=1$.

The perturbed velocity, density and magnetic field can now be
 scaled by the following slow-radius dependent variables:
\begin{equation}
{\bf{v}}=\frac{\bf{V}'}{\bar{C}_S(\xi)},\,\,\,\,
\nu=\frac{n'}{\bar{N}(\xi)},\,\,\,\,
{\bf{b}}=\frac{\bf{B}'}{\bar{B}(\xi)}.
\label{13}
\end{equation}

Inserting Eqs. (9) - (13) in Eqs. (3)-(8) and keeping leading order terms in $\epsilon$ yield the resulting dimensionless dynamical equations:
\begin{equation}
\frac{\partial v_r}{\partial \tau}-2v_\theta-\frac{1}{\beta(\xi)\bar{n}(\eta)}\frac{\partial b_r}{\partial \eta}=-\frac{1}{\beta(\xi)\bar{n}(\eta)}K\big{(}\frac{\nu}{\bar{n}(\eta)}\big{)}\frac{\partial b_r}{\partial \eta}-v_z\frac{\partial v_r}{\partial \eta},\,\,\,\,
\label{14}
\end{equation}
\begin{equation}
\frac{\partial v_\theta}{\partial \tau}
+\frac{1}{2}v_r-\frac{1}{\beta(\xi)\bar{n}(\eta)}\frac{\partial b_\theta}{\partial \eta}
=-\frac{1}{\beta(\xi)\bar{n}(\eta)}K\big{(}\frac{\nu}{\bar{n}(\eta)}\big{)}\frac{\partial b_\theta}{\partial \eta}-v_z\frac{\partial v_\theta}{\partial \eta},\,\,\,\,
\label{15}
\end{equation}
\begin{equation}
\frac{\partial b_r}{\partial \tau}-\frac{\partial v_r}{\partial \eta}
=-\frac{\partial (v_z b_r)}{\partial \eta},\,\,\,\,
\label{16}
\end{equation}
\begin{equation}
\frac{\partial b_\theta}{\partial \tau}
-\frac{\partial v_\theta}{\partial \eta}
+\frac{3}{2}b_r
=-\frac{\partial (v_z b_\theta)}{\partial \eta},\,\,\,\,
\label{17}
\end{equation}
and
\begin{equation}
\frac{\partial v_z}{\partial \tau}
+\frac{\partial}{\partial \eta}\big{(}\frac{\nu}{\bar{n}(\eta)}\big{)}
=K\big{(}
 \frac{\nu}{\bar{n}(\eta)}
 \big{)}
 \frac{\partial}{\partial \eta}\big{(}\frac{\nu}{\bar{n}(\eta)}\big{)}
-\frac{1}{2}\frac{\partial v_z^2}{\partial \eta}
-\frac{1}{2}\frac{1}{\beta(\xi)\bar{n}(\eta)}
\frac{\partial (b_\theta^2+b_r^2)}{\partial \eta} ,\,\,\,\,
\label{18}
\end{equation}
\begin{equation}
\frac{\partial \nu}{\partial \tau}
+\frac{\partial [\bar{n}(\eta)v_z]}{\partial \eta}
=-
 \frac{\partial (\nu v_z)}{\partial \eta},\,\,\,\,
\label{19}
\end{equation}
where
$$
K\big{(}
 \frac{\nu}{\bar{n}(\eta)}
 \big{)}=\frac{\nu}{\bar{n}(\eta)}\big{/}\big{(}
1+ \frac{\nu}{\bar{n}(\eta)}\big{)}.
\,\,\,\,\,\,\,\,\,\,\,\,\,\,\,\,\,\,\,\,\,\,\,\,\,\,\,\,
\,\,\,\,\,\,\,\,\,\,\,\,\,\,\,\,\,\,\,\,\,\,\,\,\,\,\,\,
\,\,\,\,\,\,\,\,\,\,\,\,\,\,\,\,\,\,\,\,\,\,\,\,\,\,\,\,
\,\,\,\,\,\,\,\,\,\,\,\,\,\,\,\,\,\,\,\,\,\,\,\,\,\,\,\,
\,\,\,\,\,\,\,\,\,\,\,\,\,\,\,\,\,\,\,\,\,\,\,\,\,\,\,\,
\,\,\,\,\,\,\,\,\,\,\,\,\,\,\,\,\,\,\,\,\,\,\,\,\,\,\,\,
\,\,\,\,\,\,\,\,\,\,\,\,\,\,\,\,\,\,\,\,\,\,\,\,\,\,\,\,
\,\,\,\,\,\,\,\,\,\,\,\,\,\,\,\,\,\,\,\,\,\,\,\,\,\,\,\,
\,\,\,\,\,\,\,\,\,\,\,\,\,\,\,\,\,\,\,\,\,\,\,\,\,\,\,\,
\,\,\,\,\,\,\,\,\,\,\,\,\,\,\,\,\,\,\,\,\,\,\,\,\,\,\,\,
\,\,\,\,\,\,\,\,\,\,\,\,\,\,\,\,\,\,\,\,\,\,\,\,\,\,\,\,
$$
In addition, the system is subject to the
 following
   boundary conditions:
\begin{equation}
b_r=b_\theta=0,\,\,\,\,\nu=0\,\,\,\,\,\mbox{for}\,\,\,\,\,\eta=\pm\infty.
\label{20}
\end{equation}
The equation for the perturbed axial magnetic field decouples from the rest of the equations and drops out from the governing system of equations. Note that the radial derivatives drop from the resulting system to leading order in $\epsilon$  without approximation of frozen radial variable, and the radial dependence enters only through the beta parameter  $\beta=\beta(\xi)$.

Equations (14)-(17)  may be reduced by eliminating $v_r$ and $v_\theta$ to the following form:
\begin{equation}
\frac{\partial^2 b_r}{\partial \tau^2}-2\frac{\partial b_\theta}{\partial \tau}
-\frac{1}{\beta(\xi)}\frac{\partial }{\partial \eta}
\big{[}\frac{1}{\bar{n}(\eta)}\frac{\partial b_r}{\partial \eta}\big{]}-3 b_r=
N_r(b_r,b_\theta,\nu,v_z),\,\,\,\,
\label{21}
\end{equation}
\begin{equation}
\frac{\partial^2 b_\theta}{\partial \tau^2}+2\frac{\partial b_r}{\partial \tau}
-\frac{1}{\beta(\xi)}\frac{\partial }{\partial \eta}
\big{[}\frac{1}{\bar{n}(\eta)}\frac{\partial b_\theta}{\partial \eta}\big{]}=
N_\theta(b_r,b_\theta,\nu,v_z),
\label{22}
\end{equation}
while relations (18)-(19) may be rewritten as  follows:
\begin{equation}
\frac{\partial^2 \nu}{\partial \tau^2}
-\frac{\partial }{\partial \eta}
\big{[}\bar{n}(\eta)\frac{\partial }{\partial \eta}
\big{(}\frac{\nu}{\bar{n}(\eta)}\big{)}
\big{]}=
N(b_r,b_\theta,\nu,v_z),
\label{23}
\end{equation}
\begin{equation}
\frac{\partial \nu}{\partial \tau}
+\frac{\partial [\bar{n}(\eta)v_z]}{\partial \eta}
=
N_z(\nu,v_z).
\label{24}
\end{equation}
Here
$$
N_r(b_r,b_\theta,\nu,v_z)=-
\frac{1}{\beta(\xi)}\frac{\partial }{\partial \eta}
\big{[}
K\big{(}\frac{\nu}{\bar{n}(\eta)}\big{)}\frac{1}{\bar{n}(\eta)}
\frac{\partial b_r}{\partial \eta}
\big{]}
-\frac{\partial^2 (v_z b_r)}{\partial \tau\partial \eta}
+2\frac{\partial (v_z b_\theta)}{\partial \eta}
-\frac{\partial }{\partial \eta}
\big{[}
v_z\frac{\partial b_r}{\partial \tau}
\big{]}
-\frac{\partial }{\partial \eta}
\big{[}
v_z\frac{\partial (v_z b_r)}{\partial \eta}
\big{]}
,\,\,\,\,\,\,\,\,\,\,\,\,\,\,\,\,\,\,\,\,\,\,\,\,\,\,\,\,\,\,\,\,
\,\,\,\,\,\,\,\,\,\,\,\,\,\,\,\,\,\,\,\,\,\,\,\,\,\,\,\,
\,\,\,\,\,\,\,\,\,\,\,\,\,\,\,\,\,\,\,\,\,\,\,\,\,\,\,\,
\,\,\,\,\,\,\,\,\,\,\,\,\,\,\,\,\,\,\,\,\,\,\,\,\,\,\,\,
$$
$$
N_\theta(b_r,b_\theta,\nu,v_z)=-
\frac{1}{\beta(\xi)}\frac{\partial }{\partial \eta}
\big{[}
K\big{(}\frac{\nu}{\bar{n}(\eta)}\big{)}\frac{1}{\bar{n}(\eta)}
\frac{\partial b_\theta}{\partial \eta}
\big{]}
-\frac{\partial^2 (v_z b_\theta)}{\partial \tau\partial \eta}
-2\frac{\partial (v_z b_r)}{\partial \eta}
-\frac{\partial }{\partial \eta}
\big{[}
v_z\frac{\partial b_\theta}{\partial \tau}
\big{]}
-\frac{\partial }{\partial \eta}
\big{[}
v_z\frac{\partial (v_z b_\theta)}{\partial \eta}
\big{]}
,\,\,\,\,\,\,\,\,\,\,\,\,\,\,\,\,\,\,\,\,\,\,\,\,\,\,\,\,\,\,\,\,
\,\,\,\,\,\,\,\,\,\,\,\,\,\,\,\,\,\,\,\,\,\,\,\,\,\,\,\,
\,\,\,\,\,\,\,\,\,\,\,\,\,\,\,\,\,\,\,\,\,\,\,\,\,\,\,\,
\,\,\,\,\,\,\,\,\,\,\,\,\,\,\,\,\,\,\,\,\,\,\,\,\,\,\,\,,\,\,\,\,
$$
$$
N(b_r,b_\theta,\nu,v_z)=
\frac{1}{2\beta(\xi)}\frac{\partial^2 (b_r^2+b_\theta^2)}{\partial \eta^2}
-\frac{\partial }{\partial \eta}
\big{[}\bar{n}(\eta)
K\big{(}\frac{\nu}{\bar{n}(\eta)}\big{)}
\frac{\partial }{\partial \eta}\big{(}\frac{\nu}{\bar{n}(\eta)}\big{)}
\big{]}
+\frac{1}{2}\frac{\partial }{\partial \eta}
\big{[}\bar{n}(\eta)\frac{\partial v_z^2}{\partial \eta}\big{]}
-\frac{\partial (\nu v_z)}{\partial \eta},
\,\,\,\,
\,\,\,\,\,\,\,\,\,\,\,\,\,\,\,\,\,\,\,\,\,\,\,\,\,\,\,\,\,\,\,\,
\,\,\,\,\,\,\,\,\,\,\,\,\,\,\,\,\,\,\,\,\,\,\,\,\,\,\,\,
\,\,\,\,\,\,\,\,\,\,\,\,\,\,\,\,\,\,\,\,\,\,\,\,\,\,\,\,
\,\,\,\,\,\,\,\,\,\,\,\,\,\,\,\,\,\,\,\,\,\,\,\,\,\,\,\,,\,\,\,\,
$$
\begin{equation}
N_z(\nu,v_z)=-\frac{\partial (\nu v_z)}{\partial \eta}.
\label{25}
\end{equation}
As will be seen in the next section the form of Eqs. (21)-(24) is convenient for the description of the dynamical evolution of small perturbations.

\section{LINEAR PROBLEMS FOR THE AC AND MS MODES.}
Linearizing system (20)-(25) about the steady-state that is described in Eqs. (10) - (11) leads to the decoupling of the full system into two subsystems. As is evident from Eqs. (21)-(22) and (23) - (24)
 by setting the right hand sides to zero, the two decoupled systems describe the non-interacting AC and MS modes, which may be presented as follows:
\begin{equation}
f(\xi,\eta,\tau)=A\exp(-i\omega\tau)\hat{f}(\xi,\eta),\,\,\,\,
g(\xi,\eta,\tau)=H\exp(-i\omega\tau)\hat{g}(\xi,\eta),\,\,\,\,
\label{26}
\end{equation}
where $f$ and $g$ stands for any of the variables that characterize the AC and MS modes, respectively,  $\omega$ and $A$, $H$ are the complex eigenfrequency and constant amplitudes.

\subsection{The linear problem for the AC modes.}

The linearized subsystem that describes the AC modes may be cast into a fourth order differential equation for either  $\hat{v}_r$, $\hat{v}_\theta$   or  $\hat{b}_r$, $\hat{b}_\theta$:
\begin{equation}
\frac{1}{\bar{n}(\eta)}
\frac{d^2 }{d \eta^2}
\big{(}
\frac{1}{\bar{n}(\eta)}
\frac{d^2 \hat{v}_{r,\theta} }{d \eta^2}
\big{)}+
(2\omega^2+3)\beta\frac{1}{\bar{n}(\eta)}
\frac{d^2 \hat{v}_{r,\theta} }{d \eta^2}
+\omega^2(\omega^2-1)\beta^2 \hat{v}_{r,\theta}=0,\,\,\,\,
\label{27}
\end{equation}
\begin{equation}
\frac{d }{d \eta}
\big{[}
\frac{1}{\bar{n}(\eta)}
\frac{d^2 }{d \eta^2}
\big{(}
\frac{1}{\bar{n}(\eta)}
\frac{d  \hat{b}_{r,\theta} }{d \eta}
\big{)}
\big{]}
+(2\omega^2+3)\beta
\frac{d }{d \eta}
\big{(}
\frac{1}{\bar{n}(\eta)}
\frac{d  \hat{b}_{r,\theta} }{d \eta}
\big{)}
+\omega^2(\omega^2-1)\beta^2  \hat{b}_{r,\theta}=0
\label{28}
\end{equation}
that is subject to the corresponding boundary conditions
\begin{equation}
 \hat{b}_{r,\theta}=-\frac{1 }{i \omega}\frac{d \hat{v}_{r,\theta} }{d \eta}=0\,\,\,\mbox{for}\,\,\eta=\pm \infty.
\label{29}
\end{equation}

To find analytical solutions of the above boundary-value problem, the Gaussian density
  $ \bar{n}(\eta)=\exp(-\eta^2/2)$ is replaced by the following hyperbolic distribution:
  \begin{equation}
 \bar{n}(\eta)=a\,\mbox{sech}^2(b\eta).
\label{30}
\end{equation}
The two shape parameters $a$ and $b$ are determined to fit the disc's total mass and moment of inertia. Since a and b come out to be closed to unity ($a=\sqrt{\pi^3/24}\approx 1.1$ and $b=\sqrt{\pi^2/12}\approx 0.9$), both will be taken for convenience to be equal to unity, $a\approx b\approx 1$. Defining now the new independent variable
\begin{equation}
 \zeta= \mbox{tanh}\eta,
\label{31}
\end{equation}
the modified density axial distribution is given by:	
\begin{equation}
\bar{n}(\zeta)=1- \zeta^2,
\label{32}
\end{equation}
and the eigenvalue problem (27), (29) may be cast in the following form:
\begin{equation}
(L+
K
)(L+
K
)\hat{v}_{r,\theta}=0,\,\,\,\,
 \hat{b}_{r,\theta}\equiv-\frac{1 }{i \omega}(1- \zeta^2)\frac{d \hat{v}_{r,\theta} }{d \zeta}=0
 \,\,\,\mbox{for}\,\,\,\zeta=\pm 1,
\label{33}
\end{equation}
where $L$  is the Legendre operator:
\begin{equation}
L=\frac{d  }{d \zeta}\big{[}(1- \zeta^2)\frac{d  }{d \zeta}\big{]},\,\,\,\,\,\,
K=\beta\frac{3+2\omega^2+l\sqrt{9+16\omega^2}}{2}.
\label{34}
\end{equation}
It is easy to show that the boundary conditions are satisfied
for the discrete values $K$
\begin{equation}
K_k
=k(k+1),
\,\,\,\,k=1,2,\dots.
\label{35}
\end{equation}
This yields the following dispersion relation for the AC modes:
\begin{equation}
\omega^4\beta^2-\beta(\beta+6\beta_k)\omega^2+9\beta_k(\beta_k-\beta)=0,\,\,\,\
\beta_k=\frac{k(k+1)}{3},
\label{36}
\end{equation}
where $\beta_k$  is the critical plasma beta of the $k$-th  AC mode. Thus, the radial velocity is governed by Legendre' eigenvalue problem, and the eigenfunctions of the linear AC problem are now given by:
\begin{equation}
i \hat{v}_{r,k,l}=\omega_{k,l}P_k(\zeta),\,\,\,
\hat{v}_{\theta,k,l}=\frac{3\beta_k-\beta\omega^2_{k,l}}{2\beta}P_k(\zeta),\,\,\,\,\,
 \hat{b}_{r,k,l}=(1- \zeta^2)\frac{d P_k }{d \zeta},\,\,\,
 i \hat{b}_{\theta,k,l}=\frac{3(\beta-\beta_k)+\beta\omega^2_{k,l}}{2\beta\omega_{k,l}}(1- \zeta^2)\frac{d P_k }{d \zeta},\,\,\,\,\,
\label{37}
\end{equation}
where
$ k=1,2,...\dots\,\,\,l=\pm 1$;
 $P_k(\zeta)$  is the Legendre polynomial of order $k$; $\omega_{k,l}$ are the eigenvalues for  slow ($l=-1$) and fast ($l=+1$) AC modes:
\begin{equation}
\omega_{k,l}=\sqrt{
\frac{\beta+6\beta_k+l\sqrt{(\beta+6\beta_k)^2-36\beta_k(\beta_k-\beta)}}{2\beta}
},\,\,\,\,\,\,\,\,\,\,\,\,\,\,\,\,
k=1,2,\dots, l=\pm 1.
\label{38}
\end{equation}
Note that while the perturbed in-plane magnetic components do indeed satisfy the
 zero
 boundary conditions, the in-plane perturbed velocities tend to infinite values as  $\zeta\to\pm1$. This however is physically plausible as the axial mass flux is indeed zero for  $\zeta\to\pm1$.
 %
Note also the  vanishing of the  hydrodynamic stresses  at $\zeta\to\pm1$.
 %
The frequencies of the fast AC mode ($l=+1$) are real for arbitrary $\beta$  while the eigenvalues of the slow AC mode ($l=-1$) are imaginary for  $\beta>\beta_k$. The slow AC modes  give rise to the familiar MRI  when the threshold  $\beta=\beta_k$, is crossed. Thus, for  $\beta_k<\beta<\beta_{k+1}$, there are exactly $k$ unstable magnetorotational modes in the system.

{\it{Eigenvalues and eigenfunctions of the AC modes for small super-criticality of the first MRI mode.}}
The weakly nonlinear analysis to be unfolded below focuses on the MRI eigenmode ($k=1, l=-1)$
 that  slightly exceeds the instability threshold $\beta=\beta_1$.
 It is therefore constructive to obtain asymptotic expressions for the eigenvalues as well for the eigenfunctions of the AC  modes in the vicinity of the threshold $\beta=\beta_1$.  In that regime a small supercriticality parameter $\delta=(\beta-\beta_1)/\beta_1$  is defined that relates the growth rate of the background MRI mode with the distance from the instability threshold:
\begin{equation}
\gamma\equiv \Im{(\omega_{1,-1})}=3
\sqrt{
\frac{\delta}{7}}+O(\delta),\,\,\,\,\,\Re{(\omega_{1,-1})}\equiv0,\,\,\,(\mbox{i.e.} \,\, \gamma\sim\sqrt{\delta}),
\label{39}
\end{equation}
where  the symbols $\Re$ and $\Im$ indicate the real and imagine parts.

All other eigenvalues of the AC modes  in the vicinity of $\beta=\beta_1$ are real and given by
\begin{equation}
\omega_{k,l}=\sqrt{
\frac{1+6\hat{\beta}_k+l\sqrt{(1+6\hat{\beta}_k)^2-36\hat{\beta}_k(\hat{\beta}_k-1)}}{2}
}+O(\gamma^2),\,\,\,\,\,\,
(k=1,\,\,\,l=+ 1,\,\,\,
k=2,3,\dots, l=\pm 1).
\label{40}
\end{equation}
Here
\begin{equation}
\hat{\beta}_k=\frac{\beta_k}{\beta_1},\,\,\,\,\,
\,\,\,\,\,\,\,\,\,\,\,\,\,\,\,\,\,\,
(\hat{\beta}_1=1).
\label{41}
\end{equation}
Typical eigenvalues for  fast and slow  AC modes calculated   for  $k=1,2,3,4$ at $\beta=\beta_1$ are presented in Table 1.

Turning now to the eigenfunctions of the  MRI mode, $k=1,l=-1$, it can be shown that $\hat{v}_{\theta,1,-1}$  and $\hat{b}_{r,1,-1}$  are of zeroth order  in $\gamma$, while  $\hat{v}_{r,1,-1}$  and $\hat{b}_{\theta,1,-1}$  are of the order of $\gamma$ in the vicinity of $\beta=\beta_1$ :
\begin{equation}
\hat{v}_{r,1,-1}=\frac{2\gamma}{3}P_k(\zeta)\sim\gamma^1,\,\,\,
\hat{v}_{\theta,1,-1}=\frac{3}{2}P_k(\zeta)\sim\gamma^0,\,\,\,\,\,
 \hat{b}_{r,1,-1}=(1- \zeta^2)\frac{d P_k }{d \zeta}\sim\gamma^0,\,\,\,
\hat{b}_{\theta,1,-1}=-\frac{4\gamma}{9}(1- \zeta^2)\frac{d P_k }{d \zeta}\sim\gamma^1.\,\,\,\,\,
\label{42}
\end{equation}
By a similar way it is easy to show that the eigenfunctions of the rest of the AC modes, i.e. both slow- ($ k>1,l=-1$) and  fast- ( $k\geq 1,l=+1$)  AC  modes are of zeroth order in $\gamma$ in the vicinity of  $\beta=\beta_1$:
$$
i\hat{v}_{r,k,l}=\omega_{k,l}P_k(\zeta)\sim\gamma^0,\,\,\,
\hat{v}_{\theta,k,l}=\frac{3\hat{\beta}_k-\omega^2_{k,l}}{2}P_k(\zeta)\sim\gamma^0,\,\,
\,\,\,\,\,\,\,\,\,\,\,\,\,\,\,\,\,\,\,\,\,\,\,\,\,\,\,\,\,\,\,\,\,\,\,\,\,\,\,\,\,\,\,\,\,\,\,\,
\,\,\,\,\,\,\,\,\,\,\,\,\,\,\,\,\,\,\,\,\,\,\,\,\,\,\,\,\,\,\,\,\,\,\,\,\,\,\,\,\,\,\,\,\,\,\,\,
\,\,\,\,\,\,\,\,\,\,\,\,\,\,\,\,\,\,\,\,\,\,\,\,\,\,\,\,\,\,\,\,\,\,\,\,\,\,\,\,\,\,\,\,\,\,\,\,
\,\,\,\,\,\,\,\,\,\,\,\,\,\,\,\,\,\,\,\,\,\,\,\,\,\,\,\,\,\,\,\,\,\,\,\,\,\,\,\,\,\,\,\,\,\,\,\,
$$
\begin{equation}
 \hat{b}_{r,k,l}=(1- \zeta^2)\frac{d P_k }{d \zeta}\sim\gamma^0,\,\,\,
i\omega_{k,l}\hat{b}_{\theta,k,l}=\frac{3(1-\hat{\beta}_k)+\omega^2_{k,l}}{2}(1- \zeta^2)\frac{d P_k }{d \zeta}\sim\gamma^0.
\label{43}
\end{equation}

To leading order in the small supercriticality of the MRI mode, the latter satisfies a reduced second-order boundary-value problem that reflects its two-fold degeneracy at  $\beta=\beta_1$
\begin{equation}
 \hat{L}^{(0)}\hat{b}^{(0)}_{r}=0,\,\,\,\hat{b}^{(0)}_{r}(\pm1)=0,
\label{44}
\end{equation}
while the rest of the AC modes satisfy a fourth order linear eigenvalue problem at  $\beta=\beta_1$
\begin{equation}
 \hat{L}^{(1)}\hat{b}^{(1)}_{r}=0,\,\,\,\hat{b}^{(1)}_{r}(\pm1)=0.
\label{45}
\end{equation}
Here for further convenience the superscripts 0 denote the $k=1, l=-1$ slow AC mode while the superscript 1 denotes the rest of the AC modes :
\begin{equation}
 \hat{L}^{(0)}\equiv \bar{n}(\zeta)\frac{d^2}{d\zeta^2}+3\beta_1 ,\,\,\,
 \hat{L}^{(1)}\equiv
 \bar{n}(\zeta) \frac{d^2}{d\zeta^2}
 \big{[}
  \bar{n}(\zeta) \frac{d^2}{d\zeta^2}
 \big{]}
 +(2\omega^{(1)2}+3)\beta_1 \bar{n}(\zeta) \frac{d^2}{d\zeta^2}
 +\omega^{(1)2}(\omega^{(1)2}-1)\beta_1^2 ,
\label{46}
\end{equation}
\begin{equation}
\omega^{(1)}=\omega_{k,l},\,\,\,
 \hat{b}^{(0)}_{r}=\hat{b}_{r,k,l}\,\,\,\,(k=1,\,l=-1),\,\,\,
 \hat{b}^{(1)}_{r}=\hat{b}_{r,k,l}\,\,\,\,(k=1,\,l=+1;\,\,\,k>1,\,l=\pm1).
\label{47}
\end{equation}

 \begin{table*}
 \centering
 \begin{minipage}{140mm}
\caption{
Critical plasma beta,  frequency and  mismatch, $\beta_{k}$, $\omega_{k,l}$  and $\Delta_{k,l}$ (the difference between the frequency of an instead of the AC mode, $\omega_{k,l}$, and the closest Gaussian MS mode frequency, $\omega_M=\sqrt{M}$, $M=1,2, \cdots$),   for slow/fast-AC modes ($k=1,l=+1$; $k>1,l=\pm1$) and for
   the background  MRI mode ($k=1,l=-1$).
      }
      \begin{tabular}{@{}ccccc@{}}
$k$ & $1$ & $2$ & $3$ & $4$
    \\
        \hline
      $\beta_k$
       & $2/3$ & $2$   &   $4$     & $20/3$
                     \\
         \hline
      $\hat{\beta}_k$
       & $1$ & $3$   & $6$     & $10$
                      \\
           \hline
        $\omega_{k,-1}$
       & $0$ & $\sqrt{\frac{19-\sqrt{145}}{2}}$
       & $\sqrt{10}$    & $\sqrt{\frac{61-\sqrt{481}}{2}}$
                          \\
                   \hline
              $\Delta_{k,-1}$
                   &
                   $0$
                    &
                    $0.071$
                    &
                    $0$
                     &
                     $-0.012$
                   \\
                    \hline
        $\omega_{k,+1}$
       & $\sqrt{7}$ & $\sqrt{\frac{19+\sqrt{145}}{2}}$
         & $\sqrt{27}$    & $\sqrt{\frac{61+\sqrt{481}}{2}}$
               \\
          \hline
        $\Delta_{k,+1}$
       &
       $0$
       &
       $-0.015$
         &
         $0$
         &
         $0.006$
                \\
\end{tabular}
\end{minipage}
\end{table*}

\subsection{The linear stability problem for the MS modes.}
Replacing the axial Gaussian distribution of the equilibrium number density by an hyperbolic profile (see above Eqs. (30)- (32)) allows to solve analytically also the eigenvalue problem for MS modes  and to obtain analytical expressions for their eigenfunctions. The resulting MS eigenmodes have been studied and found to be stable for all real values of the frequency, $\omega$ [\cite{Liverts et al. 2012a}, \cite{Liverts et al. 2012b}]. In addition, it has been proved that the corresponding spectrum of the MS mode is continuous. Thus, two linearly independent solutions of the MS eigenvalue problem (e. g. $\hat{\nu}_{+1}$  and $\hat{\nu}_{-1}$  exist for arbitrary real values of the frequency   except of $\omega^2=0$  and  $\omega^2=1$  for which the Wronskian, $W=\hat{\nu}_{+1}d\hat{\nu}_{-1}/d\zeta-\hat{\nu}_{-1}d\hat{\nu}_{+1}/d\zeta$, is identically zero. For that case, namely, $\omega^2=0$ , the eigen-solution in (43) describes a perturbation of the steady-state number density with $\hat{\nu}=1-\zeta^2$ and zero perturbation of axial velocity. That solution may be included in the unperturbed steady-state solution that was defined up to an arbitrary amplitude factor $N$ that depends on the radial variable.

 Inserting anzatz (26) into the linearized Eqs. (21) - (25) leads to the following boundary-value problem for the MS mode for Gaussian and hyperbolic (with  $a\approx b\approx 1$)cases:
\begin{equation}
\hat{N}\big{[}
    \frac{\hat{\nu}}{\bar{n}(\zeta)}
    \big{]}\equiv
\frac{d}{d\zeta}
\big{[}
\bar{n}^2(\zeta)\frac{d}{d\zeta}
    \big{(}
    \frac{\hat{\nu}}{\bar{n}(\zeta)}
    \big{)}
\big{]}
+\hat{\omega}^2
\frac{\hat{\nu}}{\bar{n}(\zeta)},\,\,\,\,
\hat{\nu}=0,\,\,\,\,\mbox{for}\,\,\,\,\zeta=\pm1,
\label{48}
\end{equation}
where  $\zeta=\sqrt{2/\pi}\int_0^\eta \bar{n}(\eta)d\eta$ is the new axial variable, and for Gaussian and hyperbolic cases, respectively:
\begin{equation}
\hat{\omega}=\omega\sqrt{\pi/2},\,\,\,\,\bar{n}(\eta)=\exp(-\eta^2/2)
\label{49}
\end{equation}
and
\begin{equation}
\hat{\omega}=\omega,\,\,\,\,\bar{n}(\eta)=1-\mbox{tanh}^2\eta.
\label{50}
\end{equation}
 Equation (48) has two linearly independent real solutions  for $\omega>1$ [\cite{Liverts et al. 2012a}, \cite{Liverts et al. 2012b}]. Expressed  such that one of the linearly independent solutions is odd while the other is even in $\zeta$ they have the form:
\begin{equation}
\frac{\hat{\nu}_{+1}(\zeta)}{\sqrt{\bar{n}(\zeta)}}=
\cos\big{(} \frac{\sigma}{2} \log\frac{1-\zeta}{1+\zeta} \big{)}
-\frac{\zeta}{\sigma}\sin\big{(} \frac{\sigma}{2} \log\frac{1-\zeta}{1+\zeta} \big{)},
\label{51}
\end{equation}
\begin{equation}
\frac{\hat{\nu}_{-1}(\zeta)}{\sqrt{\bar{n}(\zeta)}}=
\frac{\sigma}{\sigma^2+1}\sin\big{(} \frac{\sigma}{2} \log\frac{1-\zeta}{1+\zeta} \big{)}
+\frac{\zeta}{\sigma^2+1}\cos\big{(} \frac{\sigma}{2} \log\frac{1-\zeta}{1+\zeta} \big{)},
\label{52}
\end{equation}
where  $\sigma=\sqrt{\omega^2-1}$; $\nu_{+1}$  and $\nu_{-1}$   are scaled such that
\begin{equation}
\hat{\nu}_{+1}(0)=1,\,\,\,\frac{d\nu_{+1}}{d\zeta}(0)=0,\,\,\,
\label{53}
\end{equation}
\begin{equation}
\hat{\nu}_{-1}(0)=1,\,\,\,\frac{d\nu_{-1}}{d\zeta}(0)=1.\,\,\,
\label{54}
\end{equation}

At this point it is important to notice that unlike the case of the AC modes
 it is commonly accepted that replacing the Gaussian axial distribution of the number density
 by the hyperbolic one  alters the nature of the MS spectrum.
Thus, while as is commonly accepted  the spectrum that is obtained from
a Gaussian profile (named Gaussian spectrum hereafter) is discrete with eigenvalues
that are given by $\omega_{M}^2=M$ and Hermite polynomials of order $M$, ($M=1,2,3\cdots$)
as corresponding axial velocity  eigenfunctions
[\cite{Okazaki et al. 1987}, see also  \cite{Lubow and Pringle 1993}, \cite{Shtemler et al. 2011}], notice the change
the hyperbolic number density profile gives rise to continuous spectrum
(named hyperbolic spectrum hereafter) [\cite{Liverts et al. 2012a}, \cite{Liverts et al. 2012b}].
 None the less as Figs. 1 and 2 indicate that the  typical eigenfunctions of both spectra
 are very close to each other at common eigenvalues, namely $\omega_M^2=M, (M=1,2,3\cdots$).



%

\begin{figure}
\vbox{
\includegraphics[scale=0.9]{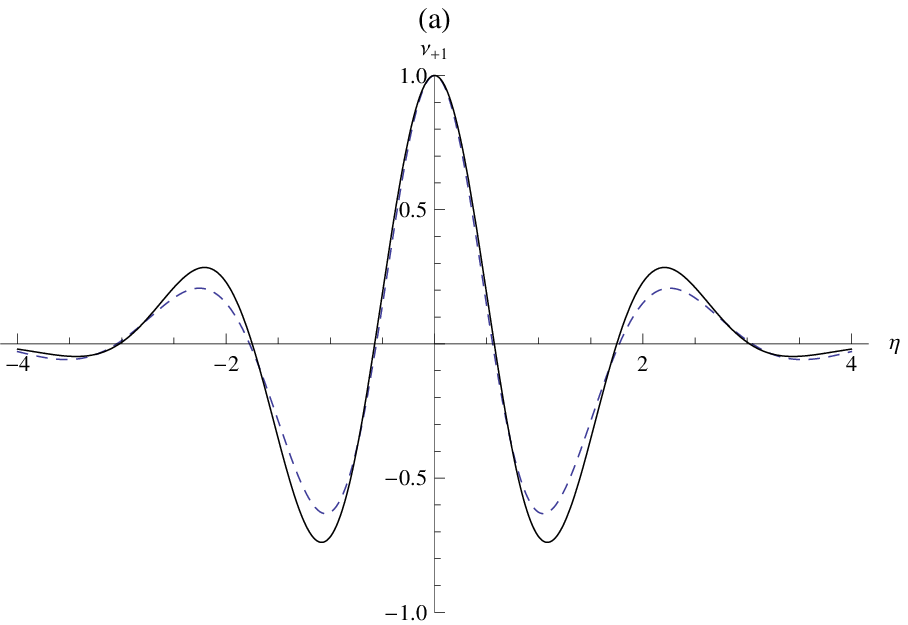}\\
\includegraphics[scale=0.9]{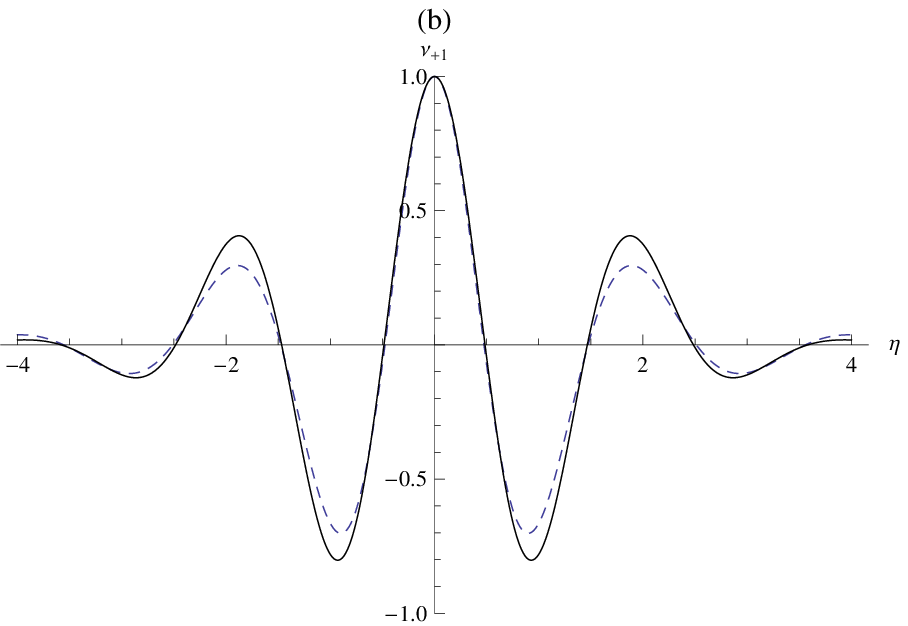}
\includegraphics[scale=0.9]{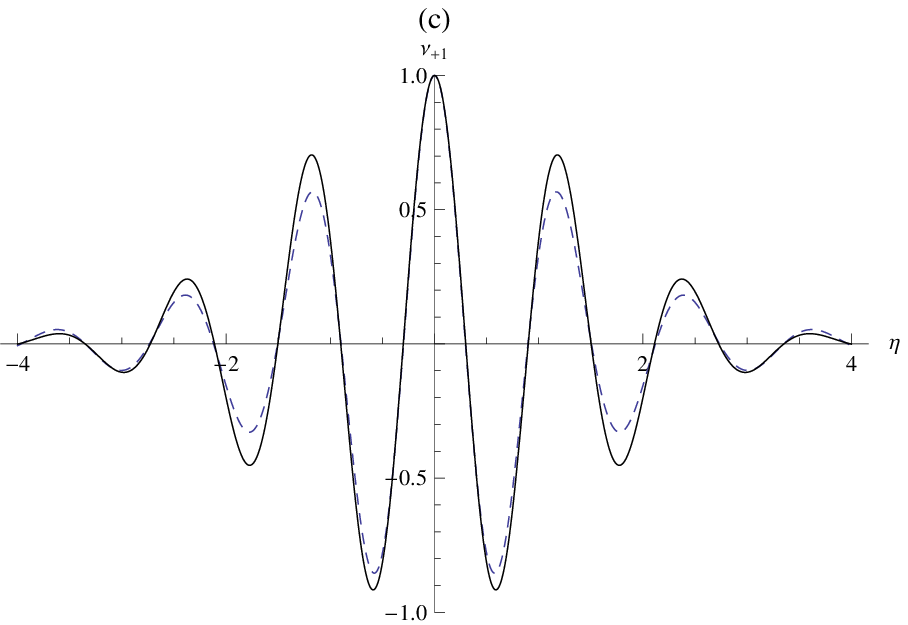}
}
\caption
{
Even MS-eigenfuctions for perturbed number density
$\hat{\nu}_{+1}$    vs $\eta$
$\big{(}
 \hat{\nu}_{+1}(0)=1,\,\,\,\,\frac{d \hat{\nu}_{+1}}{d\eta}(0)=0
 \big{)}.$
 Solid and dashed lines depict Gaussian and hyperbolic equilibrium number density, respectively.
	$(a)\,\, k=1,\,\,l=+1,\,\, \omega_{1,1}=\sqrt{7};$ 	
$(b)\,\, k=3,\,\, l=-1,\,\,\omega_{3,-1}=\sqrt{10};$
$	(c) \,\, k=3,\,\,l=+1,\,\, \omega_{3,+1}=\sqrt{27}.$
 }
\label{fig1}
\end{figure}

\begin{figure}
\vbox{
\includegraphics[scale=0.9]{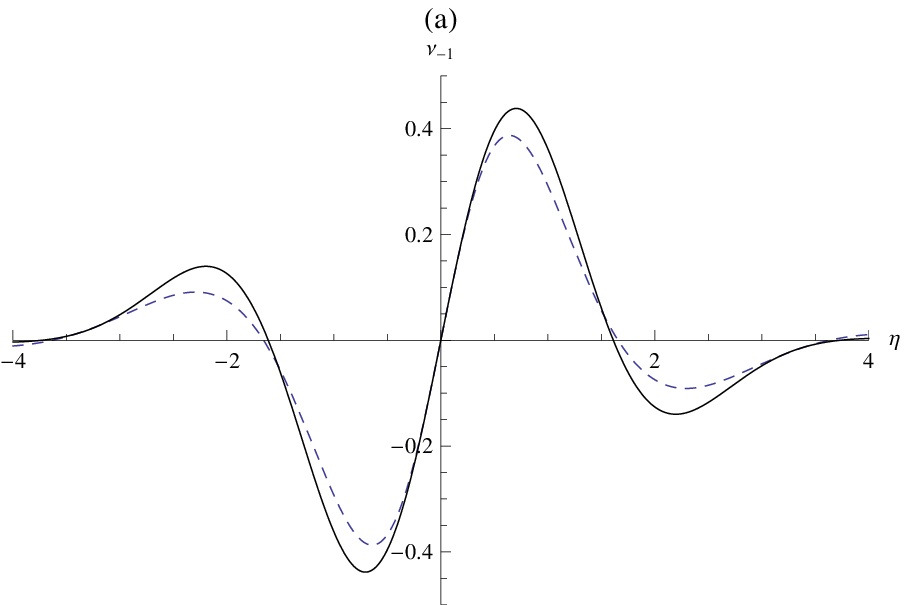}
\includegraphics[scale=0.9]{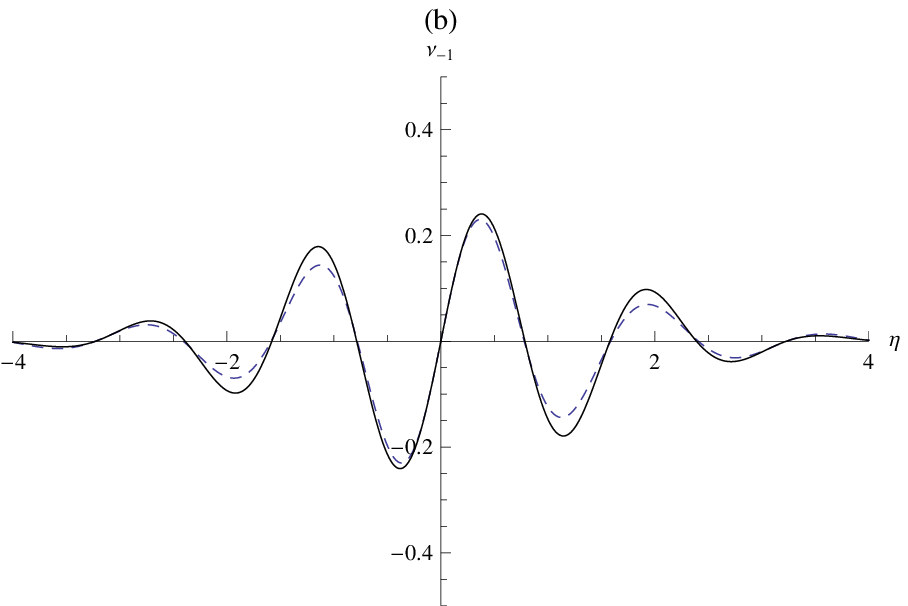}\\
\includegraphics[scale=0.9]{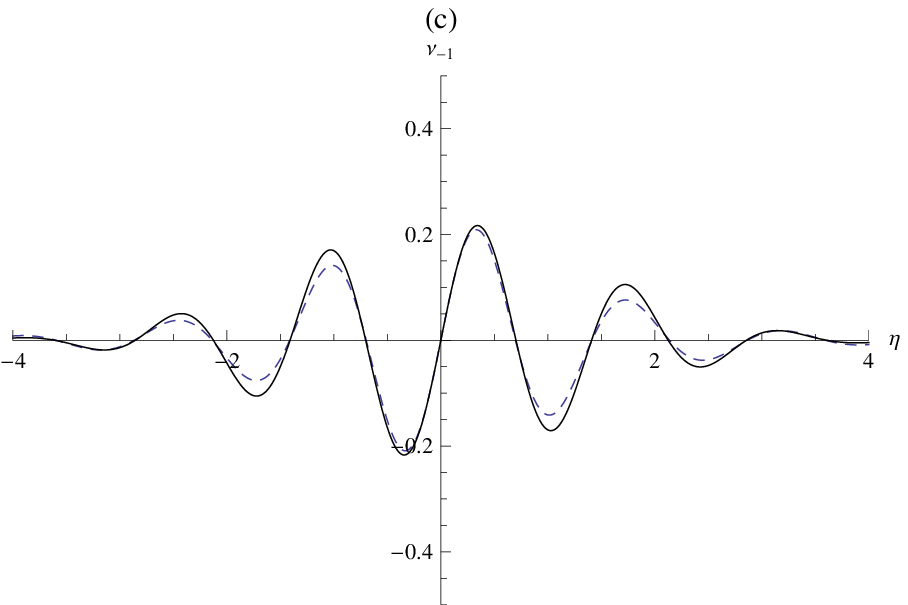}
\includegraphics[scale=0.9]{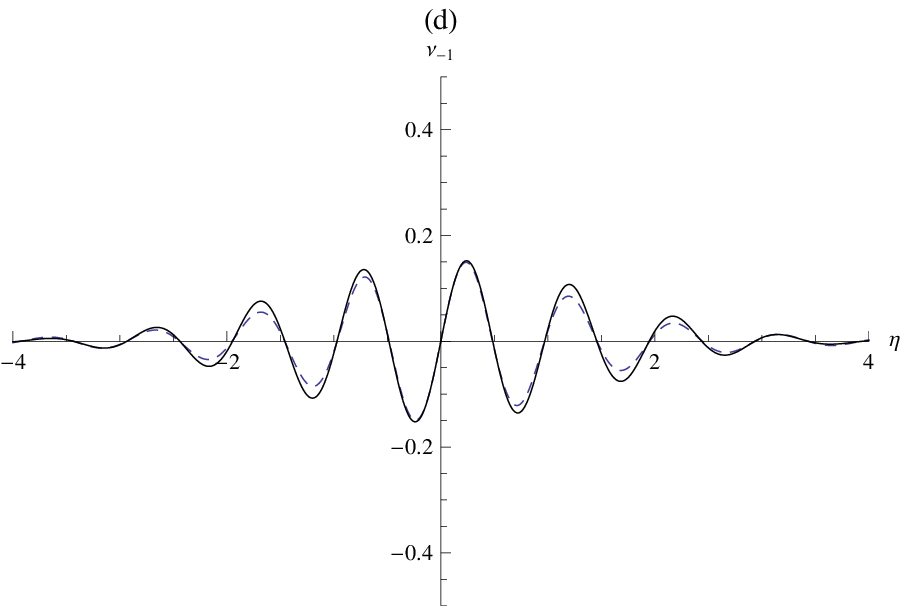}
}
\caption
{
Odd  MS-eigenfuctions for perturbed number density
$\hat{\nu}_{-1}$    vs $\eta$
$\big{(}
 \hat{\nu}_{-1}(0)=0,\,\,\,\,\frac{d \hat{\nu}_{-1}}{d\eta}(0)=1
 \big{)}$.
 Solid and dashed lines depict Gaussian and hyperbolic equilibrium number density, respectively.
	$(a)\,\, k=2,\,\,l=-1,$  $\omega_{2,-1}=\sqrt{\frac{19-\sqrt{145}}{2}};$ 	
$(b)\,\, k=2,\,\, l=+1$, $\omega_{2,1}=\sqrt{\frac{19+\sqrt{145}}{2}};$ 	
$	(c) \,\, k=4,\,\,l=-1$, $\omega_{4,-1}=\sqrt{\frac{61-\sqrt{481}}{2}};$
$	(d) \,\, k=4,\,\,l=+1$, $\omega_{4,+1}=\sqrt{\frac{61+\sqrt{481}}{2}}.$
 }
\label{fig2}
\end{figure}

\section{NON-RESONANT INTERACTION OF MRI AND MRI-DRIVEN MS MODES.}
In this section the non-resonant forcing of a MS mode by a background MRI mode [\cite{Liverts et al. 2012a}, \cite{Liverts et al. 2012b}] is reviewed  for completeness. The hierarchy of the weakly nonlinear systems is derived below in details and solved by expanding all dependent variables in terms of the small growth rate of the background MRI mode, $\gamma$.  As the real part of the MRI eigenfrequency  is zero, the original MRI mode is labeled by superscript zero as a zero harmonic. In order to describe the dynamics of that interaction a  slow time is introduced as a first step:
\begin{equation}
\tilde{\tau}=\gamma \tau\sim\gamma^0.
\label{55}
\end{equation}
Next, it is plausible to assume the following form for the MRI and for the MS variables:
$$
f^{(0)}(\zeta,\tau)=A^{(0)}(\tilde{\tau})\hat{f}^{(0)}(\zeta)+
A^{(0)}(\tilde{\tau})H^{(0)}(\tilde{\tau})\hat{f}^{(0,0)}(\zeta)+\dots,\,\,\,
\,\,\,\,
\,\,\,\,\,\,\,\,\,\,\,\,\,\,\,\,\,\,\,\,\,\,\,\,\,\,\,\,\,\,\,\,
\,\,\,\,\,\,\,\,\,\,\,\,\,\,\,\,\,\,\,\,\,\,\,\,\,\,\,\,
\,\,\,\,\,\,\,\,\,\,\,\,\,\,\,\,\,\,\,\,\,\,\,\,\,\,\,\,
\,\,\,\,\,\,\,\,\,\,\,\,\,\,\,\,\,\,\,\,\,\,\,\,\,\,\,\,,\,\,\,\,
\,\,\,\,\,\,\,\,\,\,\,\,\,\,\,\,\,\,\,\,\,\,\,\,\,\,\,\,
\,\,\,\,\,\,\,\,\,\,\,\,\,\,\,\,\,\,\,\,\,\,\,\,\,\,\,\,,\,\,\,\,
$$
\begin{equation}
g^{(0)}(\zeta,\tau)=H^{(0)}(\tilde{\tau})\hat{g}^{(0,0)}(\zeta)+\dots,\,\,\,(H^{(0)}=A^{(0)2}),
\label{56}
\end{equation}
where $f^{(0)}$  that characterize the MRI mode stands for any of $b_r^{(0)}$, $b_\theta^{(0)}$ and $v_r^{(0)}$, $v_\theta^{(0)}$, while  $g^{(0)}$ is any one from the $v_z^{(0)}$,  $\nu^{(0)}$ for the  MS modes; $\hat{f}^{(0)}(\zeta)$  denotes the MRI eigenmode; $\hat{f}^{(0,0)}(\zeta)$  denotes the back reaction of the MRI-driven MS mode on the original MRI mode;  $\hat{g}^{(0,0)}(\zeta)$ arises due to the MRI-driven MS mode, while zero MS eigenmode is absent in (56) since it doesn't belong to a  spectrum of the MS mode;
 $A^{(0)}(\tilde{\tau})$    is the amplitude of the  MRI mode, and  $H^{(0)}(\tilde{\tau})$  denotes the amplitude of the MRI-driven MS mode.

 In order to derive the equation that governs the dynamical evolution of $A^{(0)}(\tilde{\tau})$  it is first assumed that the amplitude of the  MRI is of the order of  $\gamma$, i.e. $A^{(0)}(\tilde{\tau})\sim\gamma$. Adopting therefore the estimates that are described in Eq. (42)  implies the following scalings for the MRI mode:
\begin{equation}
\{v_r^{(0)},v_\theta^{(0)}\}\sim A^{(0)}\{\hat{v}_r^{(0)},\hat{v}_\theta^{(0)}\}\sim
\{\gamma^2,\gamma^1\},\,\,\,
\{b_r^{(0)},b_\theta^{(0)}\}\sim A^{(0)}\{\hat{b}_r^{(0)},\hat{b}_\theta^{(0)}\}\sim
\{\gamma^1,\gamma^2\},\,\,\,(A^{(0)}\sim\gamma),
\label{57}
\end{equation}
where  the superscript $0$ denotes the zero harmonics in the non-stretched time, $\tau$.

In accordance with relations (21) - (25), the influence of the nonlinear effects on the background MRI mode may occur only through its nonlinear interaction with the MS mode. However, since the MS modes have only non-zero eigenfrequencies, the zero-harmonic MRI mode may only interact with the forced zero-harmonics of the MRI-driven MS mode. Applying therefore estimates (55) - (57) to Eqs. (23) - (24) yields
\begin{equation}
\{\nu^{(0)},v_z^{(0)}\}\sim H^{(0)}\{\hat{\nu}^{(0)},\hat{v}_z^{(0)}\}\sim \{\gamma^2,\gamma^3\},\,\,\,(H^{(0)}=A^{(0)2}\sim \gamma^2).
\label{58}
\end{equation}
Consequently, Eqs. (23)-(24) may be written using variable $\zeta$=tanh $\eta$ in the form:
\begin{equation}
\frac{\partial}{\partial\zeta}
\big{[}\frac{\nu^{(0)}}{\bar{n}(\zeta)}\big{]}=-\frac{1}{2\beta_1}\frac{1}{\bar{n}(\zeta)}
\frac{\partial b_r^{(0)2}}{\partial\zeta}+O(\gamma^4),
\label{59}
\end{equation}
\begin{equation}
\gamma\frac{\partial \nu^{(0)}}{\partial\tau}
+\bar{n}(\zeta)\frac{\partial[\bar{n}(\zeta)v_z^{(0)}]}{\partial\zeta}
=O(\gamma^5).
\label{60}
\end{equation}
The leading  terms in  Eqs. (59) and (60) are of  order  $\gamma^2$ and  $\gamma^3$, respectively.
Along with the boundary conditions (20) this yields the following expression for the zero harmonic of the forced  MS mode:
\begin{equation}
\frac{\nu^{(0)}(\zeta,\tau)}{\bar{n}(\zeta)}=-\frac{1}{2\beta_1}
\int_{-1}^\zeta \frac{1}{\bar{n}(\zeta)}
\frac{\partial b_r^{(0)2}}{\partial\zeta}d\zeta+O(\gamma^4)\sim A^{(0)2}\sim\gamma^2,
\label{61}
\end{equation}
\begin{equation}
v_z^{(0)}
=O(\gamma^3).
\label{62}
\end{equation}
Relation (61) is a thin-disc analogue of the relation between the perturbed pressure
 and magnetic field in a rotating isothermal axially-uniform infinite plasma [see Eq.(69.14),
 \cite{Landau and Lifshitz 1984}].  Consequently, Eqs. (21)-(22) for the zero-harmonic MRI mode, rewritten in terms of the variable $\zeta=$tanh$\eta$ and using the asymptotic estimates (55)-(60) in $\gamma$ are:
\begin{equation}
\bar{n}(\zeta)\frac{\partial^2 b_r^{(0)}}{\partial\zeta^2}
+3\beta_1 b_r^{(0)}
=-2 \gamma \beta_1\frac{\partial b_\theta^{(0)}}{\partial\tilde{\tau}}+
\gamma^2 \beta_1\frac{\partial^2 b_r^{(0)}}{\partial\tilde{\tau}^2}+
\bar{n}(\zeta)\frac{\partial }{\partial\zeta}\big{[}
 \frac{\nu^{(0)}}{\bar{n}(\zeta)}\frac{\partial b_r^{(0)}}{\partial\tilde{\zeta}}
\big{]}+O(\gamma^5),
\label{63}
\end{equation}
\begin{equation}
\bar{n}(\zeta)\frac{\partial^2 b_\theta^{(0)}}{\partial\zeta^2}
-2 \gamma \beta_1\frac{\partial b_r^{(0)}}{\partial\tilde{\tau}}
=O(\gamma^4),
\label{64}
\end{equation}
\begin{equation}
b_r^{(0)}=b_\theta^{(0)}=0\,\,\,\,\mbox{for}\,\,\,\zeta=\pm1.
\label{65}
\end{equation}
The left and right hand sides of Eq. (63) are of   order   $\gamma$ and $\gamma^3$, respectively, while the left hand side of Eq. (64) is of the order of $\gamma^2$. To lowest order in  $\gamma$, Eqs. (63)-(65) yield therefore the following relation between the in-plane components of the perturbed magnetic field:
\begin{equation}
b_\theta^{(0)}=-\frac{2}{3}\gamma\frac{\partial b_r^{(0)}}{\partial\tilde{\tau}}+O(\gamma^4).
\label{66}
\end{equation}
The left and right hand sides of Eq. (66) are of   order   $\gamma^2$. Substituting now $b_\theta^{(0)}$  from Eq. (66) into Eq. (63) yields:
\begin{equation}
\bar{n}(\zeta)\frac{\partial^2 b_r^{(0)}}{\partial\zeta^2}
+3\beta_1 b_r^{(0)}
=\frac{7}{3}
\gamma^2 \beta_1\frac{\partial^2 b_r^{(0)}}{\partial\tilde{\tau}^2}
+F_r^{(0,0)}
+O(\gamma^5),\,\,\,
b_r^{(0)}(\pm1)=0,
\label{67}
\end{equation}
where the left and right hand sides  are of   order  $\gamma$ and $\gamma^3$, respectively, and
\begin{equation}
F_r^{(0,0)}=-\frac{1}{2\beta_1}
\bar{n}(\zeta)\frac{\partial}{\partial\zeta}
\big{[}\big{(}
\int_{-1}^\zeta \frac{1}{\bar{n}(z)}\frac{\partial b_r^{(0)2}}{\partial z}dz\big{)}
\frac{\partial b_r^{(0)}}{\partial\zeta}
\big{]}
+O(\gamma^5).
\label{68}
\end{equation}

Remembering that the transition to instability occurs through a double zero root, the following  evolution equation for the amplitude of the MRI  [\cite{Liverts et al. 2012b}]:
\begin{equation}
\gamma^2
\frac{d^2 A^{(0)}}{d\tilde{\tau}^2}=
\gamma^2 A^{(0)}+ E^{(0,0)} H^{(0)} A^{(0)}
+O(\gamma^4),\,\,\,H^{(0)}=A^{(0)2},
\label{69}
\end{equation}
where $E^{(0,0)}$  is a real constant to be determined below.  Equation (69) is known as the Duffing equation. In accordance with the linear theory, initially,  $A^{(0)}(\tilde{\tau})=A^{+}\exp{(\tilde{\tau})}+A^{-}\exp{(-\tilde{\tau})}$, however, as the amplitude of the MRI-driven MS mode,  $H^{(0)}(\tilde{\tau})=A^{(0)2}(\tilde{\tau})$, grows and its back reaction becomes significant, and $A^{(0)}(\tilde{\tau})$    deviates significantly from its linear behavior.

Alternatively, transforming Eq. (69) to the scaled amplitudes $a^{(0)}=A^{(0)}/\gamma\sim  \gamma^0$, $h^{(0)}=H^{(0)}/\gamma^2\sim\gamma^0$ reduces it to the form that is independent of $\gamma$:
\begin{equation}
\frac{d^2 a^{(0)}}{d\tilde{\tau}^2}=
 a^{(0)}+ E^{(0,0)} h^{(0)} a^{(0)}
+O(\gamma),\,\,\,h^{(0)}=a^{(0)2}
\label{70}
\end{equation}
that is  subject initial conditions
\begin{equation}
a^{(0)}(0)=a^{(0)}_0,\,\,\,\frac{d a^{(0)}}{d\tilde{\tau}}=\dot{a}^{(0)}_0.
\label{71}
\end{equation}

Inserting Eq. (70) into equations for $b_r^{(0)}$  and  $b_r^{(0,0)}$    and equating terms of the same orders in $\gamma$, yields for the first threshold beta,   $\beta=\beta_1$:
\begin{equation}
\hat{L}^{(0)}\hat{b}_r^{(0)}=0,
\,\,\,\,\,\,\,\,\,\,\,\,\,\,\,\,\,\,\,\,\,\,\,\,\,\,\,\,\,\,\,\,\,\,\,\,\,\,\,\,\,\,\,\,\,\,\,\,\,\,\,\,
\,\,\,\,\,\,\,\,\,\,\hat{b}_r^{(0)}(\pm 1)=0,\,\,\,\,\,\,\,\,\,\,\,\,\,\,\,\,\,\,\,\,\,\,\,\,\,\,\,\,\,\,
(\hat{L}^{(0)}\equiv \frac{d^2 }{d\zeta^2}+\frac{3\beta_1}{\bar{n}(\zeta)}),
\label{72}
\end{equation}
\begin{equation}
\hat{L}^{(0)}\hat{b}_r^{(0,0)}=E^{(0,0)} \hat{F}^{(0)}+\hat{F}^{(0,0)},\,\,\,\,\,\,\,\,
\,\,\,\hat{b}_r^{(0,0)}(\pm 1)=0.
\label{73}
\end{equation}
The leading  terms in  Eq. (72) and (73) are of   order   $\gamma$ and $\gamma^3$, respectively,
$$
\hat{F}^{(0)}=\frac{7}{3}\beta_1\frac{\hat{b}_r^{(0)}}{\bar{n}(\zeta)},\,\,\,\,
\hat{F}^{(0,0)}=\frac{d}{d\zeta}\big{[}\frac{\hat{\nu}^{(0)}}{\bar{n}(\zeta)}\frac{d\hat{b}_r^{(0)}}{d\zeta}\big{]}
,\,\,\,
\frac{\hat{\nu}^{(0)}(\zeta)}{\bar{n}(\zeta)}=-\frac{1}{2\beta_1}\int_{-1}^\zeta \frac{1}{\bar{n}(\zeta)}\frac{d\hat{b}_r^{(0)2}}{d\zeta}d\zeta.
\,\,\,\,\,\,\,\,\,\,\,\,\,\,\,\,\,\,\,\,\,\,\,\,\,\,\,\,\,\,\,\,\,\,\,\,\,\,\,\,\,\,\,\,\,\,\,\,\,\,\,\,
\,\,\,\,\,\,\,\,\,\,\,\,\,\,\,\,\,\,\,\,\,\,\,\,\,\,\,\,\,\,\,\,\,\,\,\,\,\,\,\,\,\,\,\,\,\,\,\,\,\,\,\,
\,\,\,\,\,\,\,\,\,\,\,\,\,\,\,\,\,\,\,\,\,\,\,\,\,\,\,\,\,\,\,\,\,\,\,\,\,\,\,\,\,\,\,\,\,\,\,\,\,\,\,\,
\,\,\,\,\,\,\,\,\,\,\,\,\,\,\,\,\,\,\,\,\,\,\,\,\,\,\,\,\,\,\,\,\,\,\,\,\,\,\,\,\,\,\,\,\,\,\,\,\,\,\,\,
\,\,\,\,\,\,\,\,\,\,\,\,\,\,\,\,\,\,\,\,\,\,\,\,\,\,\,\,\,\,\,\,\,\,\,\,\,\,\,\,\,\,\,\,\,\,\,\,\,\,\,\,
$$
The self-adjoint eigenvalue problem (72) obviously describes the linear eigenfunctions near the first instability threshold (that may be expressed in terms of the Legendre polynomials (see Eqs. (42) and (43)).
%
The nonlinear coupling coefficient $E^{(0,0)}$  is determined from the solvability condition of the non-homogeneous problem (73) for $\hat{b}_r^{(0,0)}$:

\begin{equation}
E^{(0,0)} =-\int^{1}_{-1}\hat{F}^{(0,0)}(\zeta)\hat{b}_r^{(0)}(\zeta)d\zeta\big{/}
\int^{1}_{-1}\hat{F}^{(0)}(\zeta)\hat{b}_r^{(0)}(\zeta)d\zeta.
\label{74}
\end{equation}
The result is $E^{(0,0)}=-27/35\approx-0.77$, and consequently
Duffing equation (70) demonstrates how the amplitude of the MRI mode saturates by transferring its energy back and forth to the MS mode [\cite{Liverts et al. 2012b}].

\section{THREE-WAVE RESONANT INTERACTION. }
\subsection{ Scenario of the resonant interaction. }
Below, a resonant interaction   of a zero-harmonics  parent  MRI mode $(k=1, l=-1)$ with  two linearly stable daughter waves:  a MS mode ($m=\pm1$) and  one of the rest AC modes $(k=1, l=+1$ or $k>1$, $l=\pm1)$ is considered. Such interaction is investigated within the framework of a weakly nonlinear approximation that is valid for small supercriticality of the parent MRI mode.
Since in the thin disc approximation the axial variations are much bigger than the radial ones, the radial dependence of the eigenfunctions is parametrical, and the resonant relations are reduced to matching condition for the frequencies of the resonant participants: $\omega^{(0)}+\omega^{(1)}={\omega'}^{(1)}$. As defined earlier, superscript zero denotes the parent  MRI mode whose eigenvalue's real part is zero ($\omega^{(0)}=\omega_{1,-1}=0$), while superscript $1$ denotes the real and positive eigenvalues of the daughter waves  (AC  wave with $\omega^{(1)}=\omega_{k,l}$,  $k=1$, $l=+1$ or $k>1$, $l=\pm1$   and  MS  wave with ${\omega'}^{(1)}=\omega_M$):
\begin{equation}
\omega_M=\omega_{k,l}.
\label{75}
\end{equation}
Here  all frequencies $\omega_{k,l}$ are calculated for the threshold beta  of the parent  MRI mode.

For the  continuous  MS spectrum that is associated with the hyperbolic density profile, condition (75) can be satisfied for every AC mode.
For the discrete MS spectrum that is associated with the  Gaussian  density profile however  condition (75) is exactly satisfied for  some of the modes and approximately for the others. Thus, examining Table 1 if turns out that near $\beta_1$ for the fast AC modes $k=1,3$ as well as for the slow AC mode $k=3$, condition (75) is satisfied exactly. For the rest of the AC modes condition (75) is indeed satisfied approximately, such that the difference between the frequencies of those AC modes, $\omega_{k,l}$, and the closest  frequencies of MS modes, $\omega_M=\sqrt{M}$ are  small,
$\Delta_{k,l}=(\omega_{k,l}-\omega_M)/\omega_{k,l}\ll 1$ (Table 1).  Furthermore, as is also demonstrated in Table 1 the   value $\Delta_{k,l}$ (named the frequency mismatch)  tends fast to zero with increasing the axial wavenumber $k$.
The calculations presented in the current section have been carried out under the assumption of a perfect frequency match between the AC and MS modes. A discussion of a  three-wave interaction with notice change small non-zero frequency mismatch is presented in  Section 5.7.


 Recalling that a forced MS mode influences the weakly nonlinear behavior of the  parent MRI mode,  the asymptotic estimates  (42)-(43) and (55)-(60)  are adopted for the modes  that participate in the resonant configuration. This implies the following amplitude orders that consistent with the resulting nonlinear relations:
\begin{equation}
\{v_r^{(1)},v_\theta^{(1)},b_r^{(1)},b_\theta^{(1)}\}=
A^{(1)}\{\hat{v}_{r}^{(1)},\hat{v}_{\theta}^{(1)},\hat{b}_{r}^{(1)},\hat{b}_{\theta}^{(1)}\}\sim \gamma^{3/2},\,\,\,
\{\nu^{(1)},v_z^{(1)}\}=
H^{(1)}\{\hat{\nu}^{(1)},\hat{v}_{z}^{(1)}\}\sim \gamma^{3/2}
\label{76}
\end{equation}
 for the eigenfunctions of AC modes  $\{\hat{v}_{r}^{(1)},\hat{v}_{\theta}^{(1)},\hat{b}_{r}^{(1)},\hat{b}_{\theta}^{(1)}\}\equiv
\{\hat{v}_{r,k,l},\hat{v}_{\theta,k,l},\hat{b}_{r,k,l},\hat{b}_{\theta,k,l}\}$
with $k=1, l=+1$ or $k>1, l=\pm1$, as well as for the eigen-MS modes  $\{\hat{\nu}^{(1)},\hat{v}_{z}^{(1)}\}\equiv\{\hat{\nu}_m,\hat{v}_{z,m}\}$
 for the even/odd MS mode
  with $m=\pm1$.

Labeling the AC (MS) modes as axially even or odd according to the corresponding property of the perturbed radial magnetic field (perturbed density), since the parent  MRI mode ($k=1, l=-1$) is an odd function of the axial variable, the following assumption is adopted on symmetry grounds, which can be verified a'posteriori: the  parent   MRI mode may participate in the resonant interaction with  either (i) an even AC mode ($k=2, 4, \dots, l=\pm1$) and an odd MS mode ($m=-1$) or (ii) an odd AC mode ($k=1, l=+1$; $k=3,5,\dots, l=\pm1$) and an even MS mode ($m=+1$).

\subsection{ Two-time asymptotics for the resonant triads. }
For small super-criticality of the parent  MRI mode, the following fast and slow times $\bar{\tau}$ and $\tilde{\tau}$, respectively, naturally emerge:
\begin{equation}
\bar{\tau}=\omega^{(1)}\tau,\,\,\,
\tilde{\tau}=\gamma\tau,\,\,\,
\frac{\partial}{\partial \tau}=\omega^{(1)}\frac{\partial}{\partial \bar{\tau}}+
\gamma \frac{\partial}{\partial \tilde{\tau}},
\,\,\,(\gamma\ll\omega^{(1)}\sim \gamma^0).
\label{77}
\end{equation}
The solution of the weakly nonlinear problem for the modes that are involved in the three wave  resonant interaction is represented therefore as the following sum of zero, first and higher harmonics in the fast time scale $\bar{\tau}$:
\begin{equation}
\{f(\zeta,\tau),g(\zeta,\tau)\}=
\{f^{(0)}(\zeta,\tilde{\tau}),g^{(0)}(\zeta,\tilde{\tau})\}+
\big{[}\exp(-i \bar{\tau})\{f^{(1)}(\zeta,\tilde{\tau}),g^{(1)}(\zeta,\tilde{\tau})\}+\dots + c.c
\big{]}.
\label{78}
\end{equation}
Here $f$  stands for any of the variables that characterize the AC modes, namely,
 $b_r$, $ib_\theta$, $iv_r$, $v_\theta$, while $g$ describes the MS modes, namely,
 $iv_z$, and $\nu$; superscripts denote the harmonic-number in the fast time; $f^{(0)}(\zeta,\tilde{\tau}),g^{(0)}(\zeta,\tilde{\tau})$  and
$f^{(1)}(\zeta,\tilde{\tau}),g^{(1)}(\zeta,\tilde{\tau})$ are  amplitudes of the zero- and  first-harmonics, respectively; the dots represent higher-harmonics in the fast time which are also of higher order in the weakly nonlinear hierarchy.
Utilizing  the insight gained from the non-resonant case and remembering again that the MS spectrum does not include a zero eigenvalue, the following form is assumed for  all the modes that are involved in the resonant interaction:
$$
f^{(0)}(\zeta,\tilde{\tau})=A^{(0)}(\tilde{\tau})\hat{f}^{(0)}(\zeta)
+A^{(0)}(\tilde{\tau})H^{(0)}(\tilde{\tau})\hat{f}^{(0,0)}(\zeta)
+\big{[}A^{(1)*}(\tilde{\tau})H^{(1)}(\tilde{\tau})\hat{f}^{(-1,1)}(\zeta)+c.c.
\big{]}+\dots,\,\,
\,\,\,\,\,\,\,\,\,\,\,\,\,\,\,\,\,\,\,\,\,\,\,\,\,\,\,\,\,\,\,\,\,\,\,\,\,\,\,\,\,\,\,\,\,\,\,\,\,\,\,\,
\,\,\,\,\,\,\,\,\,\,\,\,\,\,\,\,\,\,\,\,\,\,\,\,\,\,\,\,\,\,\,\,\,\,\,\,\,\,\,\,\,\,\,\,\,\,\,\,\,\,\,\,
\,\,\,\,\,\,\,\,\,\,\,\,\,\,\,\,\,\,\,\,\,\,\,\,\,\,\,\,\,\,\,\,\,\,\,\,\,\,\,\,\,\,\,\,\,\,\,\,\,\,\,\,
\,\,\,\,\,\,\,\,\,\,\,\,\,\,\,\,\,\,\,\,\,\,\,\,\,\,\,\,\,\,\,\,\,\,\,\,\,\,\,\,\,\,\,\,\,\,\,\,\,\,\,\,
$$
$$
g^{(0)}(\zeta,\tilde{\tau})=H^{(0)}(\tilde{\tau})\hat{g}^{(0,0)}(\zeta),\,\,\,\,\,\,\,(H^{(0)}=A^{(0)2}),
\,\,\,\,\,\,\,\,\,\,\,
\,\,\,\,\,\,\,\,\,\,\,\,\,\,\,\,\,\,\,\,\,\,\,\,\,\,
\,\,\,\,\,\,\,\,\,\,\,\,\,\,\,\,\,\,\,\,\,\,\,\,\,\,\,\,\,\,\,\,\,\,\,\,\,\,\,\,\,\,\,\,\,\,\,\,\,\,\,\,
\,\,\,\,\,\,\,\,\,\,\,\,\,\,\,\,\,\,\,\,\,\,\,\,\,\,\,\,\,\,\,\,\,\,\,\,\,\,\,\,\,\,\,\,\,\,\,\,\,\,\,\,
\,\,\,\,\,\,\,\,\,\,\,\,\,\,\,\,\,\,\,\,\,\,\,\,\,\,\,\,\,\,\,\,\,\,\,\,\,\,\,\,\,\,\,\,\,\,\,\,\,\,\,\,
\,\,\,\,\,\,\,\,\,\,\,\,\,\,\,\,\,\,\,\,\,\,\,\,\,\,\,\,\,\,\,\,\,\,\,\,\,\,\,\,\,\,\,\,\,\,\,\,\,\,\,\,
\,\,\,\,\,\,\,\,\,\,\,\,\,\,\,\,\,\,\,\,\,\,\,\,\,\,\,\,\,\,\,\,\,\,\,\,\,\,\,\,\,\,\,\,\,\,\,\,\,\,\,\,
\,\,\,\,\,\,\,\,\,\,\,\,\,\,\,\,\,\,\,\,\,\,\,\,\,\,\,\,\,\,\,\,\,\,\,\,\,\,\,\,\,\,\,\,\,\,\,\,\,\,\,\,
$$
$$
f^{(1)}(\zeta,\tilde{\tau})=A^{(1)}(\tilde{\tau})\hat{f}^{(1)}(\zeta)
+A^{(0)}(\tilde{\tau})H^{(1)}(\tilde{\tau})\hat{f}^{(0,1)}(\zeta)+c.c.+\dots,\,\,
\,\,\,\,\,\,\,\,\,\,\,\,\,\,\,\,\,\,\,\,\,\,\,\,\,\,\,\,\,\,\,\,\,\,\,\,\,\,\,\,\,\,\,\,\,\,\,\,\,\,\,\,
\,\,\,\,\,\,\,\,\,\,\,\,\,\,\,\,\,\,\,\,\,\,\,\,\,\,\,\,\,\,\,\,\,\,\,\,\,\,\,\,\,\,\,\,\,\,\,\,\,\,\,\,
\,\,\,\,\,\,\,\,\,\,\,\,\,\,\,\,\,\,\,\,\,\,\,\,\,\,\,\,\,\,\,\,\,\,\,\,\,\,\,\,\,\,\,\,\,\,\,\,\,\,\,\,
\,\,\,\,\,\,\,\,\,\,\,\,\,\,\,\,\,\,\,\,\,\,\,\,\,\,\,\,\,\,\,\,\,\,\,\,\,\,\,\,\,\,\,\,\,\,\,\,\,\,\,\,
\,\,\,\,\,\,\,\,\,\,\,\,\,\,\,\,\,\,\,\,\,\,\,\,\,\,
$$
\begin{equation}
g^{(1)}(\zeta,\tilde{\tau})=H^{(1)}(\tilde{\tau})\hat{g}^{(1)}(\zeta)
+A^{(0)}(\tilde{\tau})A^{(1)}(\tilde{\tau})\hat{g}^{(0,1)}(\zeta)+c.c.+\dots,
\label{79}
\end{equation}
where  $f^{(0)}$, $f^{(1)}$, and $g^{(1)}$  are the known eigenfunctions of the participating eigenmodes, and $f^{(0,0)}$, $f^{(-1,1)}$, $f^{(0,1)}$ and  $g^{(0,0)}$, $g^{(0,1)}$   are yet to be determined.
 To summarize, the picture that emerges is that a triad of small eigen perturbations that satisfy the resonant condition initially coexist without mutual interaction. As one of the modes, namely the  MRI, grows in time, the resonant interaction between the triad members starts to be effective and the various modes change accordingly. In addition, the growing MRI inevitably non-resonantly excites a zero frequency forced MS mode with an amplitude $H^{(0)}(\tilde{\tau})=A^{(0)2}(\tilde{\tau})$.  The goal therefore is to find the appropriate dynamical equations for the long-time scale evolution of the amplitudes of the initial eigen perturbations, namely, $A^{(0)}(\tilde{\tau})$,  $A^{(1)}(\tilde{\tau})$, and  $H^{(1)}(\tilde{\tau})$. This is accomplished in the rest of the current section.

The physically plausible evolution equations for the amplitudes of the participating eigenmodes are given by
\begin{equation}
\gamma^2
\frac{d^2 A^{(0)}}{d\tilde{\tau}^2}=
\gamma^2 A^{(0)}+ E^{(0,0)} H^{(0)} A^{(0)}+
[E^{(-1,1)} A^{(1)*}H^{(1)} +c.c.]
+O(\gamma^4),\,\,\,
\label{80}
\end{equation}
\begin{equation}
\gamma
\frac{d A^{(1)}}{d\tilde{\tau}}=
i D^{(0,1)}  A^{(0)} H^{(1)}
+O(\gamma^{7/2}),\,\,\,
\label{81}
\end{equation}
\begin{equation}
\gamma
\frac{d H^{(1)}}{d\tilde{\tau}}=
i C^{(0,1)} A^{(0)} A^{(1)}
+O(\gamma^{7/2}),\,\,\,
\label{82}
\end{equation}
where $c.c.$ denotes complex conjugate values;
 the order of the MRI mode $A^{(0)}$  in $\gamma$ is the same as in the non-resonant case;
  the orders of $A^{(1)}$ and $H^{(1)}$  are chosen to make  the resonant terms in the amplitude relations (80)-(82) of the same order as the non-resonant  ones  in small  $\gamma$:
\begin{equation}
A^{(0)}\sim \gamma,\,\,\,H^{(0)}= A^{(0)2} \sim \gamma^2,\,\,\,
A^{(1)}\sim H^{(1)}\sim \gamma^{3/2}.\,\,\,
\label{83}
\end{equation}
These estimates are provided by the proper choice of initial conditions for $A^{(0)}$, $A^{(1)}$ and $H^{(1)}$.
The nonlinear terms with real  coefficients $E^{(0,0)}$, $ E^{(-1,1)}$,  $D^{(0,1)}$   and $C^{(0,1)}$  are introduced in the amplitude equations (80)-(82)  in order to eliminate secular terms from the corresponding nonhomogeneous
problems by satisfying the solvability conditions (see the next sub-section):
\begin{equation}
 E^{(0,0)}\sim E^{(-1,1)}\sim  D^{(0,1)}\sim  C^{(0,1)}
\sim \gamma^0.\,\,\,
\label{84}
\end{equation}
%
Introducing the  scaled amplitudes of  zeroth order in  $\gamma$ yields
\begin{equation}
a^{(0)}=\frac{A^{(0)}}{\gamma}\sim \gamma^0,\,\,\,h^{(0)}=\frac{H^{(0)}}{\gamma^2} \equiv a^{(0)2} \sim \gamma^0,\,\,\,
a^{(1)}=\frac{A^{(1)}}{A'\gamma^{3/2}}\sim \gamma^0,\,\,\,
 h^{(1)}=\frac{H^{(1)}}{H'\gamma^{3/2}}\sim \gamma^0,\,\,\,
\label{85}
\end{equation}
where $A'$ and  $H'$ are   scaling  real constants that are determined for further simplicity as follows:
$$
A'=\frac{1}{H'}=\sqrt[4]{\mid \frac{D^{(0,1)}}{C^{(0,1)} }\mid }.\,\,\,\,
\,\,\,\,\,\,\,\,\,\,\,\,\,\,\,\,\,\,\,\,\,\,\,\,\,\,
\,\,\,\,\,\,\,\,\,\,\,\,\,\,\,\,\,\,\,\,\,\,\,\,\,\,\,\,\,\,\,\,\,\,\,\,\,\,\,\,\,\,\,\,\,\,\,\,\,\,\,\,
\,\,\,\,\,\,\,\,\,\,\,\,\,\,\,\,\,\,\,\,\,\,\,\,\,\,\,\,\,\,\,\,\,\,\,\,\,\,\,\,\,\,\,\,\,\,\,\,\,\,\,\,
\,\,\,\,\,\,\,\,\,\,\,\,\,\,\,\,\,\,\,\,\,\,\,\,\,\,\,\,\,\,\,\,\,\,\,\,\,\,\,\,\,\,\,\,\,\,\,\,\,\,\,\,
\,\,\,\,\,\,\,\,\,\,\,\,\,\,\,\,\,\,\,\,\,\,\,\,\,\,\,\,\,\,\,\,\,\,\,\,\,\,\,\,\,\,\,\,\,\,\,\,\,\,\,\,
\,\,\,\,\,\,\,\,\,\,\,\,\,\,\,\,\,\,\,\,\,\,\,\,\,\,\,\,\,\,\,\,\,\,\,\,\,\,\,\,\,\,\,\,\,\,\,\,\,\,\,\,
\,\,\,\,\,\,\,\,\,\,\,\,\,\,\,\,\,\,\,\,\,\,\,\,\,\,\,\,\,\,\,\,\,\,\,\,\,\,\,\,\,\,\,\,\,\,\,\,\,\,\,\,
$$
Recasts Eqs. (80)-(82) in the following $\gamma$-free form:
\begin{equation}
\frac{d^2 a^{(0)}}{d\tilde{\tau}^2}=
a^{(0)}+ E^{(0,0)} h^{(0)} a^{(0)}+
[E^{(-1,1)} a^{(1)*}h^{(1)} +c.c.]
+O(\gamma),\,\,\,(h^{(0)} \equiv a^{(0)2}),
\label{86}
\end{equation}
\begin{equation}
\frac{d a^{(1)}}{d\tilde{\tau}}=
i \,\sigma_D \Lambda  a^{(0)} h^{(1)}
+O(\gamma),\,\,\,
\label{87}
\end{equation}
\begin{equation}
\frac{d h^{(1)}}{d\tilde{\tau}}=
i\,\sigma_C \Lambda a^{(0)} a^{(1)}
+O(\gamma),\,\,\,\,\,\,
 \label{88}
\end{equation}
where $\sigma_D=\mbox{sign}(D^{(0,1)})$, $\sigma_C= \mbox{sign}(C^{(0,1)})$,
$\Lambda =\sqrt{\mid C^{(0,1)} D^{(0,1)}\mid}$.

Equations (86) - (88) are subject initial conditions:
\begin{equation}
 a^{(0)}(0)=a^{(0)}_0,\,\,\,\frac{d a^{(0)}}{{d\tilde{\tau}}}(0)=\dot{a}^{(0)}_0,\,\,\,
 h^{(1)}(0)=h^{(1)}_0,\,\,\,
 a^{(1)}(0)=a^{(1)}_0,\,\,\,
\label{89}
\end{equation}
where $a^{(0)}_0$  is real.

\subsection{ The coupling coefficients. }

{\it{The weakly-nonlinear problem for the parent MRI mode.}}
Substituting relations (77)-(89) of the previous subsection into Eqs. (20)-(22), and equating terms of the first and cubic orders in $\gamma$, respectively, yield the following equations for zero-harmonics  in the fast time $\bar{\tau}$ of the parent MRI mode:
\begin{equation}
 \hat{L}^{(0)} \hat{b}_r^{(0)}=0,\,\,\,\hat{b}_r^{(0)}(\pm1)=0,\,\,\,\,
 \big{(} \hat{L}^{(0)} \equiv \frac{d^2}{d\zeta^2}+\frac{3\beta_1}{\bar{n}(\zeta)}
  \big{)},
  \label{90}
\end{equation}
and, separating inputs of the non-resonant and resonant  nonlinear terms,
\begin{equation}
 \hat{L}^{(0)} \hat{b}_r^{(0,0)}=E^{(0,0)}\hat{F}^{(0)}(\zeta)+\hat{F}^{(0,0)}(\zeta),\,\,\, \,\,\,\, \,\,\,\, \,\,\,\,\hat{b}_r^{(0,0)}(\pm1)=0,
 \,\,\,\, \,\,\,\, \,\,\,\,
 \label{91}
\end{equation}
\begin{equation}
 \hat{L}^{(0)} \hat{b}_r^{(-1,1)}=E^{(-1,1)}\hat{F}^{(0)}(\zeta)+\hat{F}^{(-1,1)}(\zeta),\,\,\,\hat{b}_r^{(-1,1)}(\pm1)=0.
 \label{92}
\end{equation}
Here as previously
a single superscript indicates the harmonic-number, while two superscripts indicate the harmonic-numbers of the modes that provide the input into the corresponding nonlinear term. In particular, the interaction of the non-zero harmonics of the daughters  AC and  MS modes results in zero harmonics, which nonlinearly perturbs the zero-harmonic parent MRI mode,
$$
\hat{F}^{(0)}=\frac{7}{3}\beta_1\frac{\hat{b}_r^{(0)}}{\bar{n}(\zeta)},\,\,\,\,
\hat{F}^{(0,0)}=\frac{d}{d\zeta}\big{[}\frac{\hat{\nu}^{(0)}}{\bar{n}(\zeta)}\frac{d\hat{b}_r^{(0)}}{d\zeta}\big{]}
,\,\,\,
\hat{F}^{(-1,1)}=2\frac{d}{d\zeta}\big{[}\frac{\hat{\nu}^{(1)}}{\bar{n}(\zeta)}\frac{d\hat{b}_r^{(1)}}{d\zeta}\big{]}
+2\beta_1\frac{d}{d\zeta}\big{[}\bar{n}(\zeta)
\frac{d}{d\zeta}\big{(}\frac{\hat{\nu}^{(1)}}{\bar{n}(\zeta)}\big{)}\hat{b}_r^{(1)}
\big{]}.
  \,\,\,\, \,\,\,\, \,\,\,\, \,\,\,\, \,\,\,\, \,\,\,\, \,\,\,\, \,\,\,\, \,\,\,\, \,\,\,\, \,\,\,\, \,\,\,\, \,\,\,\, \,\,\,\, \,\,\,\, \,\,\,\, \,\,\,\, \,\,\,\, \,\,\,\, \,\,\,\, \,\,\,\, \,\,\,\, \,\,\,\, \,\,\,\, \,\,\,\, \,\,\,\, \,\,\,\, \,\,\,\, \,\,\,\, \,\,\,\, \,\,\,\, \,\,\,\, \,\,\,\, \,\,\,\, \,\,\,\, \,\,\,\, \,\,\,\, \,\,\,\, \,\,\,\, \,\,\,\, \,\,\,\, \,\,\,\, \,\,\,\, \,\,\,\, \,\,\,\, \,\,\,\, \,\,\,\, \,\,\,\, \,\,\,\, \,\,\,\, \,\,\,\, \,\,\,\, \,\,\,\, \,\,\,\, \,\,\,\, \,\,\,\, \,\,\,\, \,\,\,\, \,\,\,\, \,\,\,\, \,\,\,\, \,\,\,\, \,\,\,\, \,\,\,\, \,\,\,\, \,\,\,\, \,\,\,\,
$$
The solvability conditions of the non-homogeneous problems (91) and (92) result in the following expressions for the nonlinear coupling coefficients  $E^{(0,0)}$ and $E^{(-1,1)}$:
$$
E^{(0,0)} =-\int^{1}_{-1}\hat{F}^{(0,0)}(\zeta)\hat{b}_r^{(0)}(\zeta)d\zeta\big{/}
\int^{1}_{-1}\hat{F}^{(0)}(\zeta)\hat{b}_r^{(0)}(\zeta)d\zeta,\,\,\,
  \,\,\,\, \,\,\,\, \,\,\,\, \,\,\,\, \,\,\,\, \,\,\,\, \,\,\,\, \,\,\,\, \,\,\,\, \,\,\,\, \,\,\,\, \,\,\,\, \,\,\,\, \,\,\,\, \,\,\,\, \,\,\,\, \,\,\,\, \,\,\,\, \,\,\,\, \,\,\,\, \,\,\,\, \,\,\,\, \,\,\,\, \,\,\,\, \,\,\,\, \,\,\,\, \,\,\,\, \,\,\,\, \,\,\,\, \,\,\,\, \,\,\,\, \,\,\,\, \,\,\,\, \,\,\,\, \,\,\,\, \,\,\,\, \,\,\,\, \,\,\,\, \,\,\,\, \,\,\,\, \,\,\,\, \,\,\,\, \,\,\,\, \,\,\,\, \,\,\,\, \,\,\,\, \,\,\,\, \,\,\,\, \,\,\,\, \,\,\,\, \,\,\,\, \,\,\,\, \,\,\,\, \,\,\,\, \,\,\,\, \,\,\,\, \,\,\,\, \,\,\,\, \,\,\,\, \,\,\,\, \,\,\,\, \,\,\,\, \,\,\,\, \,\,\,\, \,\,\,\, \,\,\,\, \,\,\,\,
$$
 \begin{equation}
E^{(-1,1)} =-\int^{1}_{-1}\hat{F}^{(-1,1)}(\zeta)\hat{b}_r^{(0)}(\zeta)d\zeta
\big{/}
\int^{1}_{-1}\hat{F}^{(0)}(\zeta)\hat{b}_r^{(0)}(\zeta)d\zeta.
\label{93}
\end{equation}
Obviously,  $E^{(0,0)}$ is the same as in the non-resonant case (see Eq. (74)).
Solutions for the MS - and AC - eigenmodes are determined above in Section 3:
$ \hat{b}_r^{(0)}= \hat{b}_{r,1,-1}$,
 $ \hat{b}_\theta^{(0)}= \hat{b}_{\theta,1,-1}$ is the solution of the  eigenvalue problem for the parent  MRI mode;
 $\omega^{(1)}=\omega_{k,l}$; $\hat{b}_r^{(1)}= \hat{b}_{r,k,l}$,
 $ \hat{b}_\theta^{(1)}= \hat{b}_{\theta,k,l}$    are solutions of the  eigenvalue problem for both slow and fast daughter AC modes ($k=1,l=+1$ and $k>1, l=\pm1$);  $\nu^{(1)}=\nu_{m}$ ($m=\pm1$) are  eigenfunctions of the daughter MS modes. Due to the symmetry properties in $\zeta$, the coupling coefficients in (93) have non-zero values for the eigenmodes with  $m=-1$, $k=2,4,6\dots$,  and for $m=+1$, $k=1,3,5\dots$, while the forced MS mode is determined by Eq. (59):
\begin{equation}
\frac{\hat{\nu}^{(0)}(\zeta)}{\bar{n}(\zeta)}=-\frac{1}{2\beta_1}\int_{-1}^\zeta \frac{1}{\bar{n}(\zeta)}\frac{d\hat{b}_r^{(0)2}}{d\zeta}d\zeta.
 \label{94}
\end{equation}

 {\it{The weakly-nonlinear problem for the daughter AC modes. }} Substituting now relations  (77)-(89) of the previous subsection
 into Eqs. (20)- (22),
 equating the terms of  orders  $\gamma^{3/2}$ and $\gamma^{5/2}$, and neglecting the terms of the highest order in $\gamma$, yield
 the following equations for
 the daughter AC mode
 ($\hat{b}_\theta^{(1)}= \hat{b}_{\theta,k,l}$, $k=1,l=+1$; $k>1,l=\pm1$):
\begin{equation}
 \hat{L}^{(1)} \hat{b}_\theta^{(1)}=0,\,\,\, \,\,\,\,
 \,\,\,\, \,\,\,\, \,\,\,\, \,\,\,\, \,\,\,\, \,\,\,\, \,\,\,\, \,\hat{b}_\theta^{(1)}(\pm1)=0,
 \,\,\,\, \,\,\,\, \,\,\,\, \,\,\,\, \,\,\,\, \,\,\,\,
 \big{(} \hat{L}^{(1)} \equiv\frac{d^2}{d\zeta^2}
  \big{[}
  \bar{n}(\zeta)\frac{d^2 }{d\zeta^2}
  \big{]}
  +
  \beta_1(2\omega^{(1)2}+3)\frac{d^2 }{d\zeta^2}
  + \beta_1^2\omega^{(1)2}(\omega^{(1)2}-1)\frac{1}{ \bar{n}(\zeta)} \big{)},
 \label{95}
\end{equation}
\begin{equation}
 \hat{L}^{(1)} \hat{b}_\theta^{(0,1)}=D^{(0,1)}\hat{F}^{(1)}(\zeta)+\hat{F}^{(0,1)}(\zeta),
 \,\,\,\hat{b}_\theta^{(0,1)}(\pm1)=0,
 \label{96}
\end{equation}
where
$$
\hat{F}^{(1)}(\zeta)=
\big{[}\big{(}\beta_1(\omega^{(1)2}+3)-\frac{\hat{\beta}_k}{2}\big{)}
\big{(}\omega^{(1)2}+1-3\hat{\beta}_k\big{)}+
2\beta_1\big{(}\omega^{(1)2}+3(\hat{\beta}_k-1)\big{)}\big{]}\frac{\beta_1}{\bar{n}(\zeta)}\hat{b}_r^{(1)},
\,\,\,\,
\,\,\,\, \,\,\,\, \,\,\,\, \,\,\,\, \,\,\,\, \,\,\,\, \,\,\,\, \,\,\,\, \,\,\,\, \,\,\,\, \,\,\,\,
\,\,\,\, \,\,\,\, \,\,\,\, \,\,\,\, \,\,\,\, \,\,\,\, \,\,\,\, \,\,\,\, \,\,\,\, \,\,\,\, \,\,\,\,
\,\,\,\, \,\,\,\, \,\,\,\, \,\,\,\, \,\,\,\, \,\,\,\, \,\,\,\, \,\,\,\, \,\,\,\, \,\,\,\, \,\,\,\,
\,\,\,\, \,\,\,\, \,\,\,\, \,\,\,\, \,\,\,\, \,\,\,\, \,\,\,\, \,\,\,\, \,\,\,\, \,\,\,\, \,\,\,\,
$$
$$
\hat{F}^{(0,1)}(\zeta)=2\beta_1\frac{d^2}{d\zeta^2}
\big{[}
 \bar{n}(\zeta)\frac{d (i\hat{v}_z^{(1)}\hat{b}_r^{(0)})}{d\zeta}
\big{]}
+6\beta_1^2\frac{d (i\hat{v}_z^{(1)}\hat{b}_r^{(0)})}{d\zeta}
+2\beta_1\omega^{(1)}\frac{d}{d\zeta}\big{[}\frac{\hat{\nu}^{(1)}}{\bar{n}(\zeta)}\frac{d\hat{b}_r^{(0)}}{d\zeta}
\big{]}.
\,\,\,\,
\,\,\,\, \,\,\,\, \,\,\,\, \,\,\,\, \,\,\,\, \,\,\,\, \,\,\,\, \,\,\,\, \,\,\,\, \,\,\,\, \,\,\,\,
\,\,\,\, \,\,\,\, \,\,\,\, \,\,\,\, \,\,\,\, \,\,\,\, \,\,\,\, \,\,\,\, \,\,\,\, \,\,\,\, \,\,\,\,
\,\,\,\, \,\,\,\, \,\,\,\, \,\,\,\, \,\,\,\, \,\,\,\, \,\,\,\, \,\,\,\, \,\,\,\, \,\,\,\, \,\,\,\,
\,\,\,\, \,\,\,\, \,\,\,\, \,\,\,\, \,\,\,\, \,\,\,\, \,\,\,\, \,\,\,\, \,\,\,\, \,\,\,\, \,\,\,\,
$$
The solvability condition of the non-homogeneous problem (96) yields the nonlinear coupling coefficient, $D^{(0,1)}$:
\begin{equation}
D^{(0,1)} =-\int^{1}_{-1}\hat{F}^{(0,1)}(\zeta)\hat{b}_\theta^{(1)}(\zeta)d\zeta\big{/}
\int^{1}_{-1}\hat{F}^{(1)}(\zeta)\hat{b}_\theta^{(1)}(\zeta)d\zeta.
\label{97}
\end{equation}

{\it{The weakly-nonlinear problem for the daughter MS mode. }}Applying the same procedure  to Eq. (23)-(25) results in the following weakly nonlinear equations for the daughter MS modes:
\begin{equation}
\hat{N}\big{[}\frac{\hat{\nu}^{(1)}}{\bar{n}(\zeta)}\big{]}=0,\,\,\,\, \,\,\,\, \,\,\,\, \,\,\,\,\,\,\,\, \,\,\,\, \,\,\,\, \,\,\,\,\,\,\,\, \,\,\,\, \,\,\,\, \,\,\,\,\,\,\,\,\,\,\,\,\,\,\,\,\,\,\,\,\,\,\,\,\hat{\nu}^{(1)}(\pm1)=0,
\,\,\,\,\,\,\, \,\,\,\, \,\,\,\, \,\,\,\,\,\,\,\, \,\,\,\,\,\,\,\,\,\,\,\,\,\,\,\,\,\,\,\,\,\,\,\,
 \big{(}\hat{N}\equiv\frac{d }{d\zeta}\big{[}\bar{n}^2(\zeta)\frac{d }{d\zeta}\big{]}+\omega^{(1)2} \big{)},
\label{98}
\end{equation}
\begin{equation}
\hat{N}\big{[}\frac{\hat{\nu}^{(0,1)}}{\bar{n}(\zeta)}\big{]}=C^{(0,1)}\hat{G}^{(1)}(\zeta)+\hat{G}^{(0,1)}(\zeta),\,\,\,\,
\hat{\nu}^{(0,1)}(\pm1)=0,
\label{99}
\end{equation}
where the right hand sides of Eqs. (98) and (99) are of  orders $\gamma^{3/2}$ and $\gamma^{5/2}$, respectively,
$$
\hat{G}^{(1)}(\zeta)=2\omega^{(1)}\frac{\hat{\nu}^{(1)}}{\bar{n}(\zeta)},\,\,\,\,
\hat{G}^{(0,1)}(\zeta)=-\frac{1}{\beta_1}\frac{d}{d\zeta}
\big{[}\bar{n}(\zeta)\frac{d (\hat{b}_r^{(0)}\hat{b}_r^{(1)})}{d\zeta}
\big{]}.
\,\,\,\,
\,\,\,\, \,\,\,\, \,\,\,\, \,\,\,\, \,\,\,\, \,\,\,\, \,\,\,\, \,\,\,\, \,\,\,\, \,\,\,\, \,\,\,\,
\,\,\,\, \,\,\,\, \,\,\,\, \,\,\,\, \,\,\,\, \,\,\,\, \,\,\,\, \,\,\,\, \,\,\,\, \,\,\,\, \,\,\,\,
\,\,\,\, \,\,\,\, \,\,\,\, \,\,\,\, \,\,\,\, \,\,\,\, \,\,\,\, \,\,\,\, \,\,\,\, \,\,\,\, \,\,\,\,
\,\,\,\, \,\,\,\, \,\,\,\, \,\,\,\, \,\,\,\, \,\,\,\, \,\,\,\, \,\,\,\, \,\,\,\, \,\,\,\, \,\,\,\,
$$
The solvability condition of the non-homogeneous problem (99), yields
\begin{equation}
C^{(0,1)} =-\int^{1}_{-1}\hat{G}^{(0,1)}(\zeta)\hat{\nu}^{(1)}(\zeta)d\zeta \big{/}
\int^{1}_{-1}\hat{G}^{(1)}(\zeta)\hat{\nu}^{(1)}(\zeta)d\zeta.
\label{100}
\end{equation}

\subsection{ The reduced amplitude system for resonant triads. }

 {\it{The reduced amplitude equation for the parent MRI mode.}}
 The amplitude equations (86)-(88) that describe the dynamical evolution of the interacting modes may be simplified by multiplying Eqs. (87) and (88) by $h^{(1)*}$ and $a^{(1)*}$, respectively. Summing over the resulting equations yield:
 \begin{equation}
 \frac{d }{d\tilde{\tau}}\big{[}a^{(1)}h^{(1)*}+a^{(1)*}h^{(1)}\big{]}\equiv 0
\label{101}
\end{equation}
 or, equivalently, for real  coupling  constant $E^{(-1,1)}$
 \begin{equation}
E^{(-1,1)}\big{[}a^{(1)}h^{(1)*}+a^{(1)*}h^{(1)}\big{]}=E^{(-1,1)}_0\equiv const,\,\,\,
\label{102}
\end{equation}
where $E^{(-1,1)}_0=E^{(-1,1)}\big{[}a^{(1)}_0 h^{(1)*}_0+a^{(1)*}_0h^{(1)}_0\big{]}$.

 Consequently, Eq. (86) may be rewritten as Duffing's equation for $a^{(0)}$ with a constant real forcing term $E^{(-1,1)}_0$:
\begin{equation}
\frac{d^2 a^{(0)}}{d\tilde{\tau}^2}=
 E^{(0,0)} a^{(0)3}+a^{(0)}+E^{(-1,1)}_0
,\,\,\,\,a^{(0)}(0)=a^{(0)}_0,\,\,\,
\frac{d a^{(0)}}{{d\tilde{\tau}}}(0)=\dot{a}^{(0)}_0.
\label{103}
\end{equation}
Equation (103) possess a first  integral  of the form:
\begin{equation}
\frac{1}{2}\big{(}\frac{d a^{(0)}}{d\tilde{\tau}}\big{)}^2 +U(a^{(0)})= \frac{1}{2}\dot{a}^{(0)2}_0+U(a^{(0)}_0),
\label{104}
\end{equation}
where $U(a)$ is the potential function
$$
U(a)=-\frac{1}{4}E^{(0,0)} a^{4}-\frac{1}{2}a^2-E^{(-1,1)}_0 a.
\,\,\,\,
\,\,\,\, \,\,\,\, \,\,\,\, \,\,\,\, \,\,\,\, \,\,\,\, \,\,\,\, \,\,\,\, \,\,\,\, \,\,\,\, \,\,\,\,
\,\,\,\, \,\,\,\, \,\,\,\, \,\,\,\, \,\,\,\, \,\,\,\, \,\,\,\, \,\,\,\, \,\,\,\, \,\,\,\, \,\,\,\,
\,\,\,\, \,\,\,\, \,\,\,\, \,\,\,\, \,\,\,\, \,\,\,\, \,\,\,\, \,\,\,\, \,\,\,\, \,\,\,\, \,\,\,\,
\,\,\,\, \,\,\,\, \,\,\,\, \,\,\,\, \,\,\,\, \,\,\,\, \,\,\,\, \,\,\,\, \,\,\,\, \,\,\,\, \,\,\,\,
\,\,\,\, \,\,\,\, \,\,\,\, \,\,\,\, \,\,\,\, \,\,\,\, \,\,\,\, \,\,\,\, \,\,\,\, \,\,\,\, \,\,\,\,
\,\,\,\, \,\,\,\, \,\,\,\, \,\,\,\, \,\,\,\, \,\,\,\, \,\,\,\, \,\,\,\, \,\,\,\, \,\,\,\, \,\,\,\,
$$


The effective forcing parameter $E^{(-1,1)}_0$ of the Duffing's Eq. (103) may vary within a wide range of   values due to arbitrariness of the initial data, $a_0^{(1)}$ and $h_0^{(1)}$, some typical cases are discussed below. Starting with $E^{(-1,1)}_0=0$, steady-state solutions of the unforced Duffing's equation  $a^{(0)}+E^{(0,0)} a^{(0)3}=0$, can be characterized by the following values:
\begin{equation}
a_e^{(0)}=\pm\frac{1}{\sqrt{-E^{(0,0)}}},
\label{105}
\end{equation}
which together with the obvious trivial steady-state   solution $a_e^{(0)}= 0$  constitutes   a three fixed points system. In the general case,  the range of values of $E_0^{(-1,1)}$ for which such three fixed points exist may be obtained by considering the cubic equation $ E^{(0,0)} a^{(0)3}+a^{(0)}+E^{(-1,1)}_0=0$. Thus, three fixed points exist when the absolute value of $E^{(-1,1)}_0$ is smaller than or equals to the following value:
\begin{equation}
E^{(-1,1)}_{bf}=\pm \frac{2}{\sqrt{-27E^{(0,0)}}},\,\,\,\,\,
\label{106}
\end{equation}
while only one fixed point exists otherwise.
For $E_0^{(-1,1)}= E^{(-1,1)}_{bf}$, the three fixed points  of the Duffing equation are given by (the second one is a double root):
\begin{equation}
a^{(0)}_{bf,1}=3  E^{(-1,1)}_{bf},\,\,\,\,\,
a^{(0)}_{bf,2}=- \frac{3}{2}E^{(-1,1)}_{bf}.
\label{107}
\end{equation}
 Parameters that correspond to fixed points  of the Duffing equation and the corresponding solutions are given
in Table 2.

{\it{The reduced amplitude equations for the daughter MS and AC waves and their solutions. }}
After determining the solutions of the Duffing equation (103) for $a^{(0)}(\tilde{\tau})$, Eqs. (87) and (88) may be solved explicitly by defining a new independent time variable
\begin{equation}
\tilde{\tau}^{(1)}(\tilde{\tau})=\Lambda\int_{0}^{\tilde{\tau}}
a^{(0)}(t)dt,\,\,\,\,\,\,\,\,(\Lambda =\sqrt{\mid C^{(0,1)} D^{(0,1)}\mid}).
\label{108}
\end{equation}
Consequently, the solutions for $a^{(1)}(\tilde{\tau})$ and $h^{(1)}(\tilde{\tau})$  have  the following form (for real coefficients  $C^{(0,1)}$  and $D^{(0,1)}$):
$$
 a^{(1)}(\tilde{\tau})=a^{(1)}_0 \mbox{cosh}
 \big{(}
 \sqrt{-\sigma_{CD}}\,
 \tilde{\tau}^{(1)}(\tilde{\tau})\big{)}
 + \sigma_D   \,ih^{(1)}_0 \mbox{sinh}
 \big{(}
 \sqrt{-\sigma_{CD}}\,
 \tilde{\tau}^{(1)}(\tilde{\tau}) \big{)},\,\,\,
  \,\,\,\, \,\,\,\, \,\,\,\, \,\,\,\, \,\,\,\, \,\,\,\, \,\,\,\, \,\,\,\, \,\,\,\, \,\,\,\, \,\,\,\, \,\,\,\, \,\,\,\, \,\,\,\, \,\,\,\, \,\,\,\, \,\,\,\, \,\,\,\, \,\,\,\, \,\,\,\, \,\,\,\, \,\,\,\, \,\,\,\, \,\,\,\, \,\,\,\, \,\,\,\, \,\,\,\, \,\,\,\, \,\,\,\, \,\,\,\, \,\,\,\, \,\,\,\, \,\,\,\, \,\,\,\, \,\,\,\, \,\,\,\, \,\,\,\, \,\,\,\, \,\,\,\, \,\,\,\, \,\,\,\, \,\,\,\, \,\,\,\, \,\,\,\, \,\,\,\, \,\,\,\, \,\,\,\, \,\,\,\, \,\,\,\, \,\,\,\, \,\,\,\, \,\,\,\, \,\,\,\, \,\,\,\, \,\,\,\, \,\,\,\, \,\,\,\, \,\,\,\, \,\,\,\, \,\,\,\, \,\,\,\, \,\,\,\, \,\,\,\, \,\,\,\, \,\,\,\, \,\,\,\, \,\,\,\, \,\,\,\,
 $$
\begin{equation}
 i\, h^{(1)}(\tilde{\tau})=i\,h^{(1)}_0 \mbox{cosh}
 \big{(}
 \sqrt{-\sigma_{CD}}\,
 \tilde{\tau}^{(1)}(\tilde{\tau}) \big{)}
 - \sigma_C  \,a^{(1)}_0 \mbox{sinh}
 \big{(}\sqrt{-\sigma_{CD}}\,
 \tilde{\tau}^{(1)}(\tilde{\tau}) \big{)}.\,\,
 \label{109}
\end{equation}
Solutions (109) reveal the following result: the linearly stable daughter AC and MS waves   may be nonlinearly destabilized (while the parent MRI
mode  is   saturated). A necessary and sufficient condition for that to occur is:
\begin{equation}
\sigma_{CD}=\mbox{sign}(C^{(0,1)}D^{(0,1)})<0.
 \label{110}
\end{equation}
As may be seen in Table 3, this occurs for  the slow  AC modes  and the corresponding MS modes.
These nonlinearly growing daughter modes constitute the MRDI for which an effective nonlinear growth rate, $\Gamma_{nl}$, may be defined in the following way:
\begin{equation}
\Gamma_{nl}\equiv  \gamma \, \gamma_{nl}= \gamma \, \mid \textless{}
 a^{(0)}\textgreater{}\mid \Lambda,
 \,\,\,\,\,
 \textless{}
 a^{(0)}\textgreater{}= \lim_{\tilde{\tau} \to \infty} \frac{1}{\tilde{\tau}}\int_0^{\tilde{\tau}}a^{(0)}(t)
 d t.
 \label{111}
\end{equation}
The effective  growth rate of the nonlinear instability,  $\Gamma_{nl}$,  is proportional to the long-time average  amplitude of the nonlinearly saturated parent MRI mode, and is of the same order as the growth rate  of the  MRI eigenmode, $\gamma$.

Finally it is noted that the solution (109)   satisfies the following  first-integral relation for Eqs. (87) - (88):
\begin{equation}
  a^{(1)2}(\tilde{\tau})-\sigma_{CD}\,  h^{(1)2}(\tilde{\tau})=
 a^{(1)2}_0-\sigma_{CD}\,   h^{(1)2}_0.\,\,\,
 \label{112}
\end{equation}

\subsection{Clustering of resonant triads.}
The above solution for an isolated resonant triad can be generalized for a cluster of a finite number of resonant triads, joined via a common parent MRI mode. Summing Eqs. (103) for the isolated triads results in the problem for the common  MRI mode
\begin{equation}
\frac{d^2 a^{(0)}}{d\tilde{\tau}^2}=
 E^{(0,0)} a^{(0)3}+a^{(0)}+E^{(-1,1)}_{0,\Sigma},\,\,\,\,a^{(0)}(0)=a^{(0)}_0,\,\,\,
\frac{d a^{(0)}}{{d\tilde{\tau}}}(0)=\dot{a}^{(0)}_0,
\label{113}
\end{equation}
where  the forced term in Duffing's equation depends on the initial data for all daughter AC and MS modes of the various triads that form the cluster, and may be obtained by summing the triads inputs:
\begin{equation}
E^{(-1,1)}_{0,\Sigma}\equiv E^{(-1,1)}\sum_{j}[a^{(1)}_{0,j}h^{(1)*}_{0,j}+
a^{(1)*}_{0,j}h^{(1)}_{0,j}].
\label{114}
\end{equation}
Thus the  amplitude equations for the daughter MS and AC waves for a cluster of resonant triads differs from those for the isolate  triad by the amplitude value of the   parent MRI mode. The amplitude equation for the parent MRI mode in the cluster of  triads differs from that for the isolate  triad by the value of the forced term. The forced term  is a free parameter in the present model that depends on initial data for all  daughter AC and MS modes of the triads forming the cluster.

\subsection{ Numerical examples for resonant triads.}
 The coefficient $E^{(0,0)}=-27/35\approx-0.77$  describes the non-linear interaction of the MRI- and MRI-driven MS- modes.
  According to Eq. (102) the value of the coupling coefficient  $E^{(-1,1)}$  influences the results of simulations  indirectly through  the effective force term $E^{(-1,1)}_0$  that parametrically varies with arbitrary initial data for the daughter  AC and MS waves.
 The bifurcation value of the forced term $E_{bf}^{(-1,1)}$ in Eq. (106) provides a natural scale   for an arbitrary force parameter $E^{(-1,1)}_0$
 \begin{equation}
\bar{E}_0^{(-1,1)}=\frac{E_0^{(-1,1)}}{\mid E_{bf}^{(-1,1)}\mid }.
\label{115}
\end{equation}

 \begin{table*}
 \centering
 \begin{minipage}{140mm}
\caption{
Characteristic values of coefficients of Duffing's equation for  the  MRI mode and its characteristic nonlinear equilibrium solutions.
  }
\begin{tabular}{@{}cccccccc@{}}
$E^{(0,0)}$ & $a^{(0)}_e=\pm\frac{1}{  \sqrt{-E^{(0,0)}} } $
 &   $E^{(-1,1)}_{bf}=\pm\frac{2}{  \sqrt{-27E^{(0,0)}} }$ &
 $a^{(0)}_{bf,1}=3E^{(-1,1)}_{bf} $  &  $a^{(0)}_{bf,2}=-\frac{3}{2}E^{(-1,1)}_{bf}  $
     \\
        \hline
      $-0.77$
       & $\pm1.14$ & $\pm0.44$   &   $\pm 1.32$     & $\mp 0.67$
                     \\
         \end{tabular}
\end{minipage}
\end{table*}

The possible equilibrium amplitudes of the parent MRI mode are  presented in Table 2. For instance, $a^{(0)}_{bf,n}$ $(n=1,2)$ are the  equilibrium amplitudes which correspond to bifurcation from single to  three equilibrium solutions of Duffing's equation ($\bar{E}_0^{(-1,1)}=1$), while $a_e^{(0)}\neq 0$ is the equilibrium solution of the unforced Duffing's equation (105) ($\bar{E}^{(-1,1)}_{0}=0$).  It is instructive to compare the behavior of the parent MRI mode that is governed by the forced Duffing equation with the unforced system. Solutions of unforced Duffing's equation  [\cite{Liverts et al. 2012b}], $a^{(0)}(\tilde{\tau})$ and
$\dot{a}^{(0)}(\tilde{\tau})$, may have a clearly expressed bursty behavior for sufficiently small initial amplitudes (see Fig. 3). The burst regime may be even enhanced with the growing force term in Duffing's equation (see Figs. 4 and 5). However, the MRI mode remains to be stable and in the forced resonant case.
Phase trajectories  and characteristics of growth rates  for several values of the effective force  term $\bar{E}_0^{(-1,1)}=0,1,5$ in Duffing equation are depicted in Fig 6 (a) and (b), respectively.
Ordinates in Figures 6(b) are scaled by $\Lambda$ that makes the instability characteristics of the system independent from the axial wavenumber of the daughter waves.
Note that according to Eq. (112)  the long-time asymptotic value of
$\frac{1}{\Lambda}\frac{\tilde{\tau}^{(1)}}{\tilde{\tau}}$  equals to
the scaled  growth rate of the nonlinear instability, $\gamma_{nl}$, for AC and MS modes (in Fig. 6(b) $\frac{1}{\Lambda}\tilde{\tau}^{(1)}$  is drawn).
\begin{figure}
\vbox{
\includegraphics[scale=0.9]{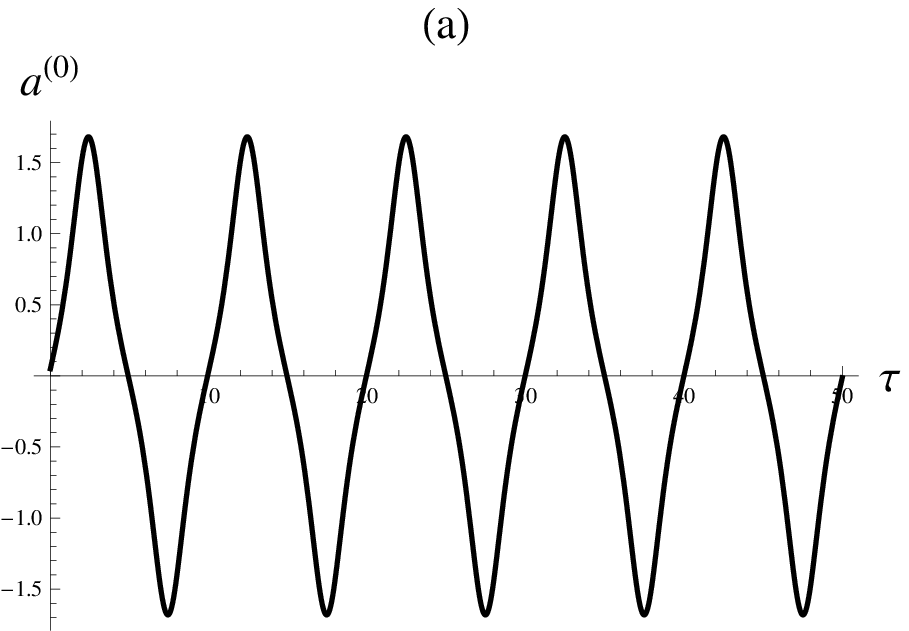}
\includegraphics[scale=0.9]{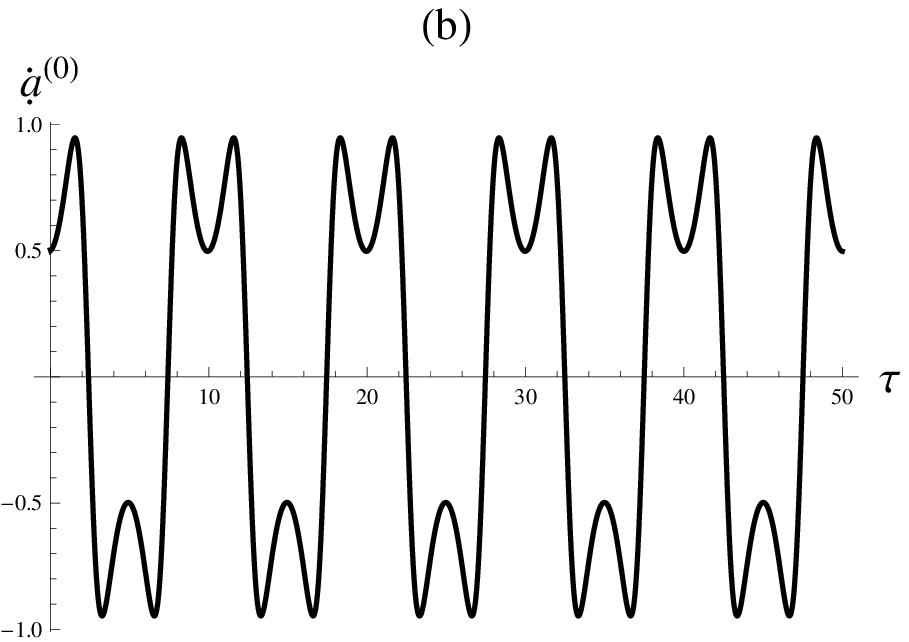}
}
\caption
{
Parent MRI mode for the unforced   system $\bar{E}_0^{(-1,1)}=0$:
  (a) $a^{(0)}(\tilde{\tau})$ ; (b) $\dot{a}^{(0)}(\tilde{\tau})$,
 $a_0^{(0)}=0.05, \dot{a}_0^{(0)}=0.5$
(tildes in  $\tilde{\tau}$ are dropped in graphics).
 }
\label{fig3}
\end{figure}

\begin{figure}
\vbox{
\includegraphics[scale=0.9]{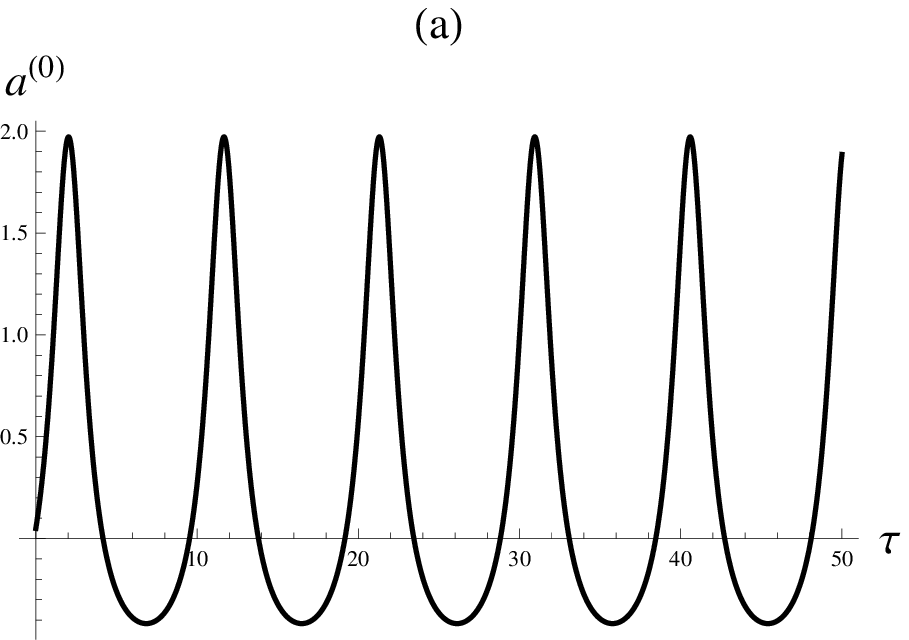}
\includegraphics[scale=0.9]{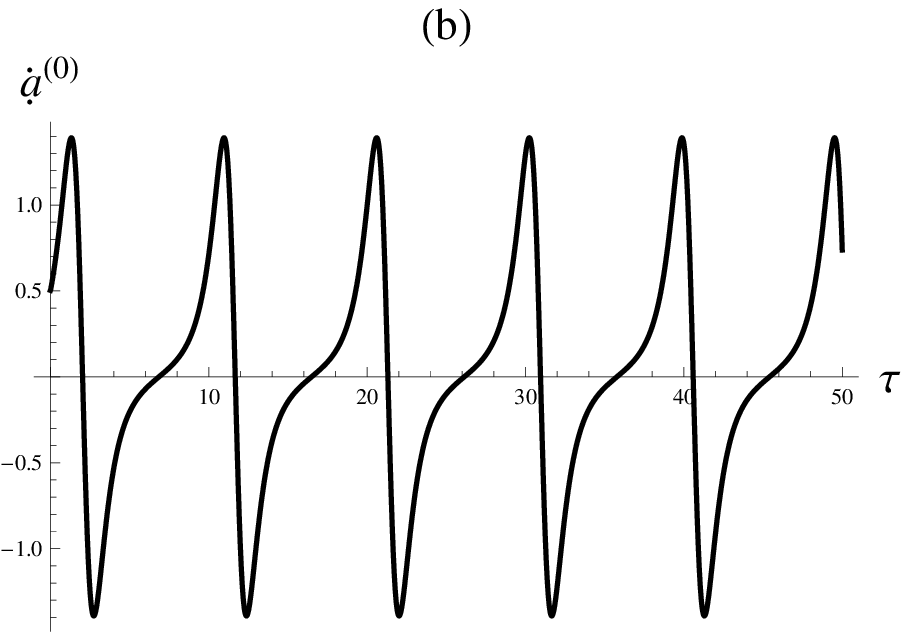}
}
\caption
{
Same as in Fig. 3 for the bifurcation value of the effective force $\bar{E}_0^{(-1,1)}=1$ ($E_{bf}^{(-1,1)} =0.44$).
    }
\label{fig4}
\end{figure}

\begin{figure}
\vbox{
\includegraphics[scale=0.9]{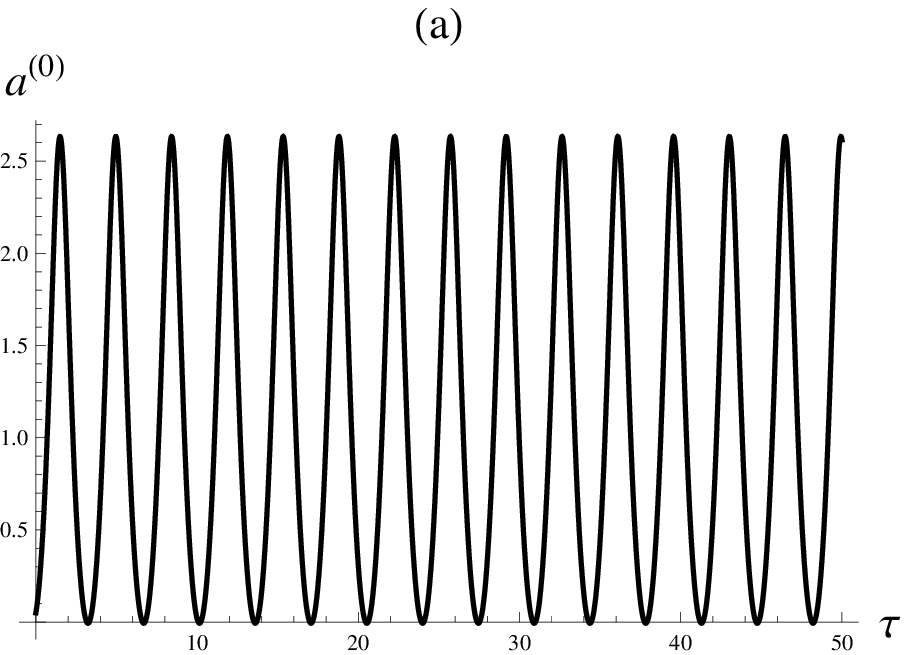}
\includegraphics[scale=0.9]{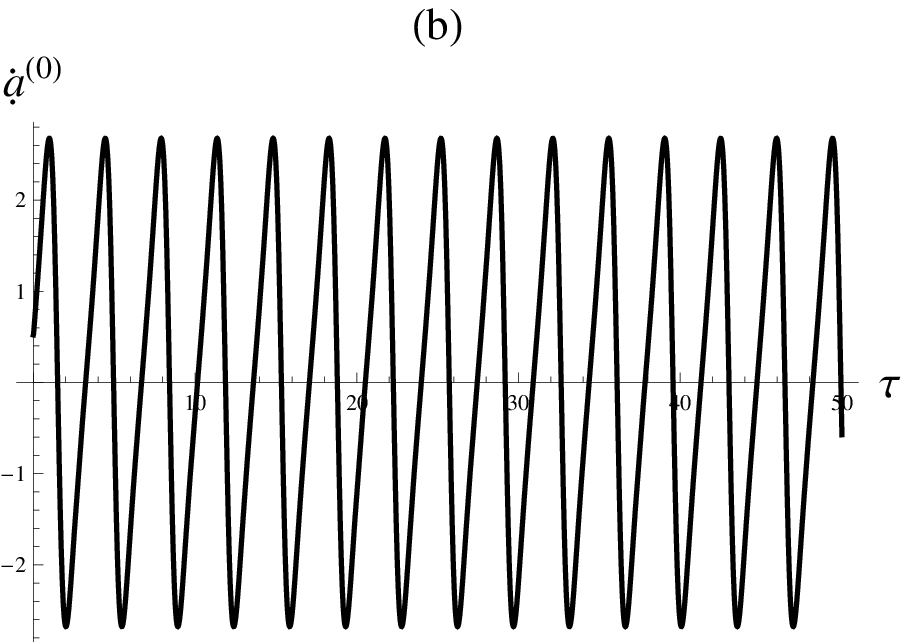}
}
\caption
{
Same as in Fig. 3 for the super-bifurcation value of the effective force    $\bar{E}_0^{(-1,1)}=5$ ($E_{bf}^{(-1,1)} =0.44$).
  }
\label{fig5}
\end{figure}

\begin{figure}
\vbox{
\includegraphics[scale=0.9]{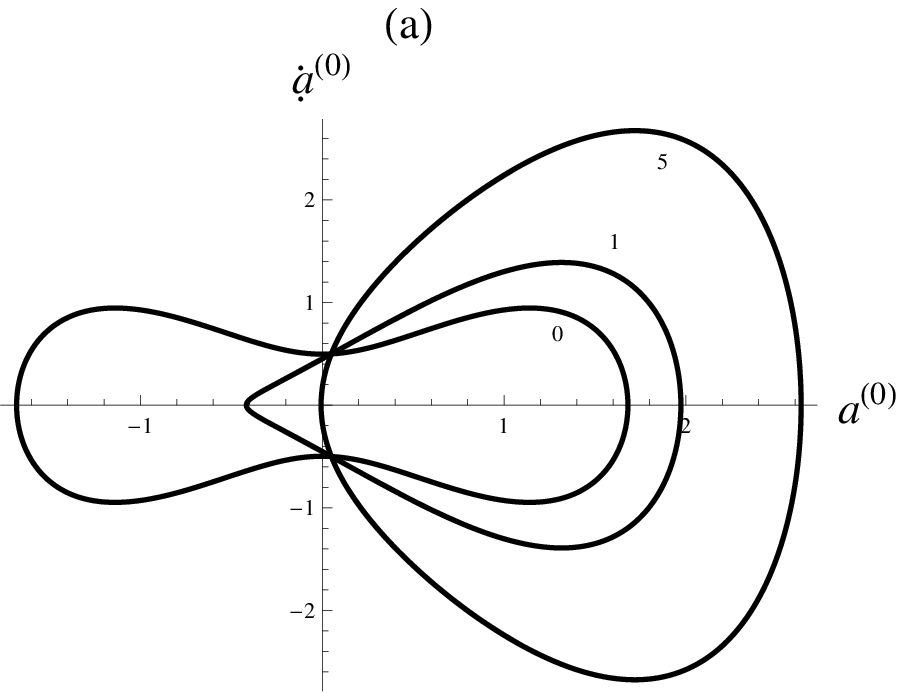}
\includegraphics[scale=0.9]{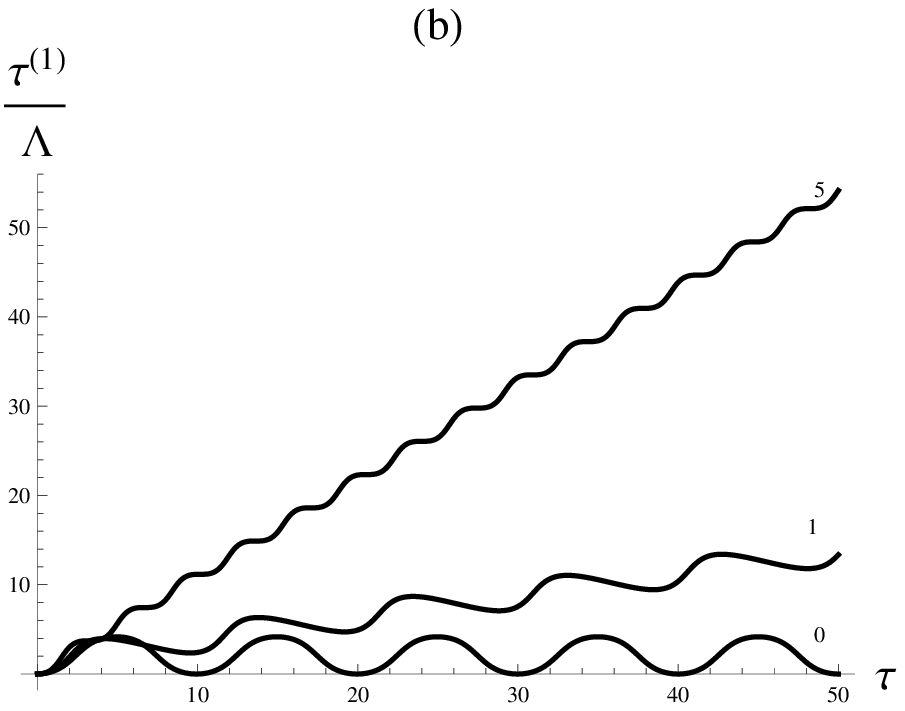}\\
}
\caption
{Characteristics of the  MRI mode (a) and of AC, MS modes (b) for several  effective-force values  $\bar{E}_0^{(-1,1)}=0,1,5$ ($E_{bf}^{(-1,1)} =0.44$). Rest of the parameters are the same as  in Fig. 3.
 (a) Phase diagram of the MRI mode, $a^{(0)}(\tilde{\tau})$ vs. $\dot{a}^{(0)}(\tilde{\tau})$;
 (b) Instability characteristics of the AC and MS modes, $\frac{\tilde{\tau}^{(1)}(\tilde{\tau})}{\Lambda}$  equals in the long-time limit  $\tilde{\tau}\to\infty$ to  $\tilde{\tau}\gamma_{nl}$, where $\gamma_{nl}=\frac{\Gamma_{nl}}{\gamma} \equiv  const$  is the growth rate of the AC and MS modes on the slow-time scale (tildes in $\tilde{\tau}^{(1)}$ and $\tilde{\tau}$ are dropped in graphics).
  }
\label{fig6}
\end{figure}

In Table 3 the real nonlinear coefficients $C^{(0,1)}$ and $D^{(0,1)}$   are given, which describe coupling of the parent MRI mode ($k=1,l=-1$) with  one of AC modes ($k=1,l=1$; $k>1,l=\pm1$) and  a  MS mode ($m=\pm1$).
The governing system for amplitudes of the  AC  and MS modes demonstrates in the limit of large
$\tilde{\tau}$, an exponential instability with a scaled constant  growth rate, $\gamma_{nl}$ for slow AC modes,   and stability  for  fast AC modes:
 (i) $\gamma_{nl}$  starts from the zero value  for the  unforced system ($\bar{E}^{(-1,1)}_0=0$),   (ii) $\gamma_{nl}$  is small  for the  sub-bifurcation value of  $\bar{E}^{(-1,1)}_{0}=1/2$  and (iii) rises up to significant values for bifurcation ($\bar{E}^{(-1,1)}_{0}=1$) and super-bifurcation ($\gamma_{nl}=5$) values of the force term in the Duffing equation  ( Table 3, see also Figs. 3-6).
 Since the scaled effective growth rate, $\gamma_{nl}=\Gamma_{nl}/\gamma$, was evaluated on the scale of slow time, $\tilde{\tau}$, on the scale of the fast time the effective growth rate is given by $\Gamma_{nl}=\gamma  \gamma_{nl}$, i.e. of the order of $\gamma$,   and the resulting instability occurs with the growth rates of the order of that for the original MRI eigenmode. As seen on Fig. 6, the long time value of $\mid \textless{}
 a^{(0)}\textgreater{}\mid \equiv const$, and according to (112)  the effective  growth rate $\Gamma_{nl}$ is constant.

 \begin{table*}
 \centering
 \begin{minipage}{140mm}
\caption{
Resonant coupling coefficients at the threshold beta of the parent MRI mode.
$\,\,\,\,\,\,\,\,\, \,\,\,\,\,\,\,\,\,\,\,\,\,\,\,\,\,\,\,\,\,\,\,\,\,\,\,\,\,\,\,\,\,\,\,\,$
 $\omega^{(1)}=\omega_{k,l}$ are the frequencies of AC modes;
  $ k\geq1$ is the axial wavenumbers of AC modes;
  $\,\,\,\,\,\,\,\,\,\,\,\,\, \,\,\,\,\,\,\,\,\,\,\,\,\,\,\,\,\,\,\,\,\,\,\,\,\,\,\,\,$
   $l=\pm1$ denotes the fast/slow AC modes  ($k=1$, $l=-1$ correspond to the parent MRI mode).
   $\,\,\,\,\,\,\,\,\,\,\,\,\,\,\,\,\,\,\,\,\,\,\,\,\,\,\,\,\,\,\,\,\,\,\,\,\,\,\,\,\,\,\,\,$
     $\Gamma_{nl}$ is the effective growth rate of the AC and MS modes at
    $\tilde{\tau}\to\infty$ ($\Gamma_{nl}=0$ for the  unforced system   $\bar{E}^{(-1,1)}_0=0$; $\gamma_{nl}\equiv \Gamma_{nl}/ \gamma $ is the scaled growth rate of order of $\gamma^0$ ).
  }
\begin{tabular}{@{}cccccccc@{}}
 & $k=1,$ &   $k=2,$ & $k=2,$ & $k=3$ & $k=3$ & $k=4$ & $k=4$ \\
& $l=1$ &   $l=-1$ & $l=1$ & $l=-1$ & $l=1$ & $l=-1$ & $l=1$
     \\
        \hline
      $\omega_{k,l}$
       & $\sqrt{7}$ & $\sqrt{\frac{19-\sqrt{145}}{2}}$   &   $\sqrt{\frac{19+\sqrt{145}}{2}}$     & $\sqrt{10}$
           & $\sqrt{27}$    & $\sqrt{\frac{61-\sqrt{481}}{2}}$    & $\sqrt{\frac{61+\sqrt{481}}{2}}$
                     \\
         \hline
      $E^{(-1,1)}$
       & $-0.14$ & $-2.16$   & $0.026$     & $3.78$
            & $0.66$     & $1.66$     & $-0.01$
                      \\
           \hline
        $C^{(0,1)}$
       & $-0.68$ & $-3.40$   &$-5.06$    & $2.01$
          & $1.81$   & $16.46$   & $13.81$
                   \\
                    \hline
        $D^{(0,1)}$
       & $-0.20$ & $94.97$    & $-0.03$    & $-4.40$    & $0.11$    & $-0.44$    & $0.011$
        \\
                    \hline
        $\gamma_{nl}$ for $\bar{E}^{(-1,1)}_0=\frac{1}{2}$
       & $0$ & $2.06$   & $0$    & $0.34$
          & $0$   & $0.31$   & $0$
        \\
                    \hline
         $\gamma_{nl}$ for $\bar{E}^{(-1,1)}_0=1$
       & $0$ & $12.4$   & $0$    & $2.05$
          & $0$   & $1.8$   & $0$
             \\
                    \hline
        $\gamma_{nl}$ for $\bar{E}^{(-1,1)}_0=5$
      & $0$ & $52.1$   & $0$    & $8.6$
          & $0$   & $7.8$   & $0$
         \\
\end{tabular}
\end{minipage}
\end{table*}

\subsection{  Resonant three-wave interaction with a small  frequency mismatch. }
As was mentioned in the previous section the continuous MS spectrum that is associated with the hyperbolic density profile enabled the frequency resonant condition to be satisfied accurately. This, however, is not the case for the Gaussian number-density profile that is commonly believed to be characterized by a discrete spectrum [\cite{Okazaki et al. 1987}, see also  \cite{Lubow and Pringle 1993}, \cite{Shtemler et al. 2011}].
Non the less, as is demonstrated above the frequency resonance condition may always be satisfied approximately up to a small mismatch that asymptotically tends to zero for increasing axial wavenumber $k$. In that case
 the relation between the frequencies of the daughter AC and MS modes may be written in the following way:
 \begin{equation}
\omega_{k,l}-\omega_M=\omega_{k,l}\gamma \tilde{\delta}_{k,l}
\label{116}
\end{equation}
where according to the principle of least degeneracy of the  problem [\cite{Van Dyke  1964}] $\tilde{\delta}_{k,l}\equiv \delta_{k,l}/\gamma$ is of order $\gamma^0$ with $\delta_{k,l}$ that is order of
$\Delta_{k,l}=(\omega_{k,l}-\omega_M)/\omega_{k,l}\ll 1$  presented in Table 1. In that case the small mismatch notice strongly influences the amplitudes of all three modes on the slow time scale $\tilde{\tau}=\gamma \tau$.
The equations for these amplitudes of the resonantly interacting modes are expected to be of the following form that generalize the system (86)-(88):
\begin{equation}
\frac{d^2 a^{(0)}}{d\tilde{\tau}^2}=
a^{(0)}+ E^{(0,0)} h^{(0)} a^{(0)}+
[E^{(-1,1)} a^{(1)*}h^{(1)}e^{-i\tilde{\tau}\tilde{\delta}_{k,l}} +c.c.]
+O(\gamma),\,\,\,(h^{(0)} \equiv a^{(0)2}),
\label{117}
\end{equation}
\begin{equation}
\frac{d a^{(1)}}{d\tilde{\tau}}=
i \,\sigma_D \Lambda  a^{(0)} h^{(1)}e^{-i\tilde{\tau}\tilde{\delta}_{k,l}}
+O(\gamma),\,\,\,
\label{118}
\end{equation}
\begin{equation}
\frac{d h^{(1)}}{d\tilde{\tau}}=
i\,\sigma_C \Lambda a^{(0)} a^{(1)}e^{+i\tilde{\tau}\tilde{\delta}_{k,l}}
+O(\gamma).\,\,\,\,\,\,
 \label{119}
\end{equation}

Indeed, \cite{Wersinger  et al. 1980a}, \cite{Wersinger  et al. 1980b} have investigated the dynamical evolution of the {\it{classical}} decay instability
with an unstable parent mode, linear damping of the daughter modes, and small frequency mismatch.
They found that for small values of linear damping of the
daughter modes the latter oscillate with exponentially growing amplitude, just as was found in the previous section for perfect frequency match.
However, as the linear damping of the daughter modes increases a richer and more complex behavior of the involved amplitudes
emerges. Thus, as the damping rate is increased the amplitudes of the daughter modes exhibit sequences of bifurcations as
well as chaotic behavior. It is expected that the solutions of system (117)-(119) will also exhibit similar complex behavior
by which the resonant three-wave interaction enhance the transport coefficient (which give rises to mode damping) that in turn make the resonant three-wave
interaction progressively more complex and chaotic thus further enhancing those transport coefficients. Such calculations are
outside the scope of the current work and will be presented in a subsequent publication.

\section{  Summary and Discussion. }

The weakly-nonlinear axisymmetric MRDI of thin Keplerian discs threaded by poloidal magnetic fields was studied in the vicinity of the threshold beta of a parent MRI mode. That instability is a direct result of the resonant coupling of three  eigenmodes of the system: the parent MRI mode and two daughter modes, one of which is a stable slow-AC modes, and the other one is a   stable MS mode. That mechanism is a natural generalization
of the fundamental decay instability discovered five decades ago for infinite, homogeneous and
stationary plasmas [\cite{Galeev  and Oraevskii 1962-1963}].
%
 Generalized for thin discs the MRDI   results in the  non-linear saturation of the parent MRI mode and the exponential growth of the daughter stable slow-AC and MS waves.  The effective growth rate of the MRDI is comparable to that  of the parent MRI eigenmode in the linear stage of its development. The MRDI occurs via  energy transfer from the MRI to  AC and MS waves as the effective growth rate of the  MRDI tends to zero with the amplitude of the MRI mode. On the other hand if the AC daughter mode is a fast, all three amplitudes remain bounded.
The weakly non-linear resonant interaction between the three modes is investigated by the method of two-scale asymptotic expansions, which reflect the slow time variation  of the parent MRI mode, and the fast oscillations of the daughter AC and MS waves.


 The  temporal evolution of the amplitudes of the resonantly interacting modes in thin discs significantly differs from its classical nonlinear predecessor [\cite{Craik 1985}]. For instance, instead of the  stable Alfv\'en wave in infinite plasma systems [\cite{Galeev  and Oraevskii 1962-1963}], the role of the parent wave is currently played by a slightly unstable first magnetorotational mode. Since the linear instability threshold is crossed at a zero eigenvalue with multiplicity two, the  amplitude of the MRI mode  is governed by an unsteady second-order forced Duffing equation. Instead of the standard first-order equation  with a quadratic  nonlinearity formed by the product of the current amplitudes of AC and MS eigenmodes, the Duffing equation contains  a cubic self-induced nonlinear term  that arises due to interaction of the parent MRI mode with the MRI-forced zero-frequency  MS perturbations, and a constant-force term of the same order that is proportional to a product  of amplitudes of the daughter AC and MS eigenmodes at the initial instant.   The resulting Duffing equation
 decouples from the   standard first-order differential equations  with quadratic nonlinear terms  for the daughter AC and MS modes.
Summarizing the principal differences of three-wave resonant instability in thin discs from infinite-plasma systems note in addition to those mentioned in Introduction: (i) the amplitude of the parent MRI mode in thin disc approximation is much larger than the amplitudes of the  two daughter  modes ($A^{(0)}\gg A^{(1)}\sim H^{(1)}$ with $A^{(0)3}\sim A^{(1)2}\sim H^{(1)2}$) instead of comparable amplitudes of all  triad's components in the infinite-plasma case ($A^{(0)}\sim A^{(1)}\sim H^{(1)}$);
 (ii) although the parent MRI mode  becomes stable in the resonant case, the daughter   AC  and MS  modes may lose their stability, as distinct from the infinite-plasma systems, in which the amplitudes of all three modes remain bounded   as they exchange energy periodically;
(iii) isolate resonant triads  in thin discs  can  form a cluster of  triads, all joined via a common parent MRI mode. In such case the amplitude of the parent MRI mode is described by an appropriate Duffing equation with a force term which depends on the initial data for all daughter AC and MS modes that form the cluster, while the amplitudes of each pairs of matched daughter waves are described as in the absence of other triads.

A resonant interaction was studied above in details for the first marginally unstable magnetorotational mode
 in a small vicinity of the  plasma beta $\beta=\beta_1<\beta_2<\beta_3 \dots$.
 %
 The  analysis presented in the current work  certainly points out to the crucial need of further investigations in order to have a better grasp of the nonlinear MHD processes  in thin rotating discs.
The exponential nonlinear growth of the slow AC and MS modes  raises some questions concerning their further behavior (either stabilization  or destabilization), as well as the system changing when the  beta exceeds the each next  threshold value, $\beta_k$, and different  MRI modes from different  clusters  start interacting with one another.
%
In particular, it is expected that the solutions of the three-mode amplitude system  will  exhibit complex  in time behavior
  when a small frequency mismatch will be taken into account.

\section*{Acknowledgments}
The current work has been supported by grant no. 180/10 of the Israel Science Foundation.

\bsp
\label{lastpage}
\end{document}